\documentclass[twocolumn]{aastex631}

\newcommand{\Msun}{{\rm M_{\odot}}}

\newcommand{\mpc}{\, {\rm Mpc}}
\newcommand{\kpc}{\, {\rm kpc}}
\newcommand{\pc}{\, {\rm pc}}
\newcommand{\K}{\, {\rm K}}
\newcommand{\kmps}{\, {\rm km \, s^{-1}}}
\newcommand{\Jykmps}{\, {\rm Jy \cdot km \, s^{-1}}}
\newcommand{\yr}{\,{\rm yr}}
\newcommand{\Myr}{\,{\rm Myr}}

\newcommand{\Mmol}{M_{\rm H_2}}
\newcommand{\Mgas}{M_{\rm gas}}
\newcommand{\Ico}{I_{\rm CO}}
\newcommand{\Sigmamol}{\Sigma_{\rm H_2}}
\newcommand{\Sigmagas}{\Sigma_{\rm gas}}
\newcommand{\Tp}{T_{\rm peak}}
\newcommand{\Tk}{T_{\rm kin}}
\newcommand{\Tb}{T_{\rm b}}

\begin{document}

\title{Diverse Molecular Structures Across The Whole Star-Forming Disk of M83: \\ High fidelity Imaging at 40pc Resolution}

\author{Jin Koda}
\affiliation{Department of Physics and Astronomy, Stony Brook University, Stony Brook, NY 11794-3800, USA}

\author{Akihiko Hirota}
\affiliation{NAOJ Chile, National Astronomical Observatory of Japan, Los Abedules 3085 Office 701, Vitacura, Santiago 763-0414, Chile}
\affiliation{Joint ALMA Observatory, Alonso de C\'ordova 3107, Vitacura, Santiago 763-0355, Chile}

\author{Fumi Egusa}
\affiliation{Institute of Astronomy, Graduate School of Science, The University of Tokyo, 2-21-1 Osawa, Mitaka, Tokyo 181-0015, Japan}

\author{Kazushi Sakamoto}
\affiliation{Academia Sinica, Institute of Astronomy and Astrophysics, Taipei 10617, Taiwan}

\author{Tsuyoshi Sawada}
\affiliation{NAOJ Chile, National Astronomical Observatory of Japan, Los Abedules 3085 Office 701, Vitacura, Santiago 763-0414, Chile}
\affiliation{Joint ALMA Observatory, Alonso de C\'ordova 3107, Vitacura, Santiago 763-0355, Chile}

\author{Mark Heyer}
\affiliation{Department of Astronomy, University of Massachusetts Amherst, 710 North Pleasant Street, Amherst, MA 01003, USA}

\author{Junichi Baba}
\affiliation{National Astronomical Observatory of Japan, 2-21-1 Osawa, Mitaka, Tokyo 181-8588, Japan.}
\affiliation{Department of Astronomy, School of Science, Graduate University for Advanced Studies (SOKENDAI), 2-21-1 Osawa, Mitaka, Tokyo, 181-1855 Japan}

\author{Samuel Boissier}
\affiliation{Aix Marseille Univ., CNRS, CNES, Laboratoire d'Astrophysique de Marseille, Marseille, France}

\author{Daniela Calzetti}
\affiliation{Department of Astronomy, University of Massachusetts Amherst, 710 North Pleasant Street, Amherst, MA 01003, USA}

\author{Jennifer Donovan Meyer}
\affiliation{National Radio Astronomy Observatory, 520 Edgemont Rd, Charlottesville, VA 22903, USA}

\author{Bruce G. Elmegreen}
\affiliation{IBM Research Division, T. J. Watson Research Center, 1101 Kitchawan Road, Yorktown Heights, NY 10598, USA}

\author{Armando Gil de Paz}
\affiliation{Departamento de F\'{\i}sica de la Tierra y Astrof\'{\i}sica, Facultad de CC$.$ F\'{\i}sicas, Universidad Complutense de Madrid, 28040, Madrid, Spain}
\affiliation{Instituto de F\'{\i}sica de Part\'{\i}culas y del Cosmos IPARCOS, Facultad de CC$.$ F\'{\i}sicas, Universidad Complutense de Madrid, 28040 Madrid, Spain}

\author{Nanase Harada}
\affiliation{National Astronomical Observatory of Japan, 2-21-1 Osawa, Mitaka, Tokyo 181-8588, Japan.}
\affiliation{Department of Astronomy, School of Science, Graduate University for Advanced Studies (SOKENDAI), 2-21-1 Osawa, Mitaka, Tokyo, 181-1855 Japan}

\author{Luis C. Ho}
\affiliation{Kavli Institute for Astronomy and Astrophysics, Peking University, Beijing 100871, China}
\affiliation{Department of Astronomy, School of Physics, Peking University, Beijing 100871, China}

\author{Masato I.N. Kobayashi}
\affiliation{National Astronomical Observatory of Japan, 2-21-1 Osawa, Mitaka, Tokyo 181-8588, Japan.}

\author{Nario Kuno}
\affiliation{Division of Physics, Faculty of Pure and Applied Sciences, University of Tsukuba, 1-1-1 Tennodai, Tsukuba, Ibaraki 305-8577, Japan}
\affiliation{Tomonaga Center for the History of the Universe (TCHoU), University of Tsukuba, 1-1-1 Tennodai, Tsukuba, Ibaraki 305-8577, Japan}

\author{Amanda M Lee}
\affiliation{Department of Physics and Astronomy, Stony Brook University, Stony Brook, NY 11794-3800, USA}

\affiliation{Department of Astronomy, University of Massachusetts Amherst, 710 North Pleasant Street, Amherst, MA 01003, USA}

\author{Barry F. Madore}
\affiliation{The Observatories, Carnegie Institution for Science, 813 Santa Barbara Street, Pasadena CA 91101 USA}

\author{Fumiya Maeda}
\affiliation{Institute of Astronomy, Graduate School of Science, The University of Tokyo, 2-21-1 Osawa, Mitaka, Tokyo 181-0015, Japan}

\author{Sergio Mart\'in}
\affiliation{European Southern Observatory, Alonso de C\'ordova, 3107, Vitacura, Santiago 763-0355, Chile}
\affiliation{Joint ALMA Observatory, Alonso de C\'ordova 3107, Vitacura, Santiago 763-0355, Chile}

\author{Kazuyuki Muraoka}
\affiliation{Department of Physics, Graduate School of Science, Osaka Metropolitan University, 1-1 Gakuen-cho, Naka-ku, Sakai, Osaka 599-8531, Japan}

\author{Kouichiro Nakanishi}
\affiliation{National Astronomical Observatory of Japan, Mitaka, Tokyo 181-8588, Japan}
\affiliation{Department of Astronomy, School of Science, Graduate University for Advanced Studies (SOKENDAI), 2-21-1 Osawa, Mitaka, Tokyo, 181-1855 Japan}

\author{Sachiko Onodera}
\affiliation{School of Science and Engineering, Meisei University, Hino, Tokyo 191-8506, Japan}

\author{Jorge L. Pineda}
\affiliation{Jet Propulsion Laboratory, California Institute of Technology, 4800 Oak Grove Drive, Pasadena, CA 91109-8099, USA} %\\

\author{Nick Scoville}
\affiliation{Division of Physics and Astronomy, California Institute of Technology, Pasadena, CA 91125, USA}

\author{Yoshimasa Watanabe}
\affiliation{Materials Science and Engineering, College of Engineering, Shibaura Institute of Technology, 3-7-5 Toyosu, Koto-ku, Tokyo 135-8548, Japan}

\begin{abstract}
We present ALMA imaging of molecular gas across the full star-forming disk of the barred spiral galaxy M83 in CO($J$=1-0).
We jointly deconvolve the data from ALMA's 12m, 7m, and Total Power arrays using the MIRIAD package.
The data have a mass sensitivity and resolution of $10^4~\Msun$ ($3\sigma$) and 40~pc -- sufficient to detect and resolve a typical molecular cloud in the Milky Way with a mass and diameter of $4\times 10^5~\Msun$ and $40\pc$, respectively.
The full disk coverage shows that the characteristics of molecular gas change radially from the center to outer disk, with the locally measured brightness temperature, velocity dispersion,
and integrated intensity (surface density) decreasing outward.
The molecular gas distribution shows coherent large-scale structures in the inner part, including the central concentration, offset ridges along the bar, and prominent molecular spiral arms. 
However, while the arms are still present in the outer disk, they appear less spatially coherent, and even flocculent.
Massive filamentary gas concentrations are abundant even in the interarm regions.
Building up these structures in the interarm regions would require a very long time ($\gtrsim 100$~Myr).
Instead, they must have formed within stellar spiral arms and been released into the interarm regions.
For such structures to survive through the dynamical processes, the lifetimes of these structures and their constituent molecules and molecular clouds must be long ($\gtrsim 100\Myr$).
These interarm structures host little or no star formation traced by H$\alpha$.
The new map also shows extended CO emission, which likely represents an ensemble of unresolved molecular clouds.

\end{abstract}


\section{Introduction} \label{sec:intro}

Molecular gas is a key link
between star formation and galaxy evolution across cosmic time \citep{Tacconi:2020wt}.
In particular, molecular clouds ($10^{4-7}~\Msun$)
host virtually all star formation in the local Universe, and
their formation and evolution are the first crucial steps leading to star formation.
Galactic dynamics around stellar bars and spiral arms stir the gas and stimulate
the formation and evolution of molecular clouds.
Subsequent star formation and feedback into the gas are two of
the most important factors in galaxy growth and quenching, respectively.
These processes leave imprints on the distribution and physical conditions
of molecular gas and clouds over a galactic disk.
To gain a more complete understanding of these processes,
it is essential to map the full population of molecular gas and clouds over a galaxy.

Until recently, molecular gas studies over large spiral galaxies
have been limited by sensitivity and resolution.
The previous CO(1-0) studies of M51 had a cloud detection limit of $\gtrsim 10^5~\Msun$,
and left significant CO emission undetected or unresolved below this limit \citep{Koda:2009wd, Koda:2011nx, Schinnerer:2013aa, Pety:2013aa}.
The molecular clouds with mass $\gtrsim 10^5~\Msun$ in the Milky Way (MW) contain
only $\sim$half of the total molecular mass \citep{Scoville:1987vo, Heyer:2015qy},
raising a natural question as to how the other \textit{hidden} half of molecular gas, presumably within smaller clouds with $\lesssim 10^5~\Msun$, contributes
to gas evolution and star formation.
A cloud-scale resolution of $40\pc$, the typical diameter of Galactic molecular clouds \citep{Scoville:1987vo},
is necessary for mapping the full cloud population and resolving cloud properties
\citep{Donovan-Meyer:2012fk, Schinnerer:2013aa, Donovan-Meyer:2013uq, Colombo:2014aa, Rosolowsky:2021aa}.

Even with ALMA, deep observations covering large galactic disks
are expensive in time.
The recent ALMA large survey of nearby galaxies \citep[PHANGS; ][]{Leroy:2021ab}
set the target sensitivity to $\sim 10^5~\Msun$ and the resolution to $\sim 100\pc$,
and observed only the inner parts of galactic disks.
In addition, many large surveys with ALMA rely on
the excited transition of CO(2-1), instead of CO(1-0),
for a higher observational efficiency (by a factor of $\sim8$).
This compromise is inevitable for large surveys,
but could introduce a bias toward warmer and/or denser gas.
Indeed, environmental variations of the CO 2-1/1-0 line ratio
have been detected in galaxies \citep{Sakamoto:1997ys, Koda:2012lr, Koda:2020aa, Yajima:2021aa, den-Brok:2021aa}.
We therefore need CO(1-0) observations to take a full census of molecular gas in galactic disks.

Currently, the only extragalactic studies of lower-mass clouds ($\sim 10^4~\Msun$) throughout a disk are limited to 
the closest dwarfs and dwarf-like spirals
\citep[LMC, SMC, M33, and NGC300; e.g., ][]{Mizuno:2001qq, Fukui:2008tl, Rosolowsky:2003mw, Tosaki:2011fk, Gratier:2012wb, Faesi:2014aa}.
However, the gas evolution in these galaxies appears substantially different from that
in large spiral galaxies, such as the MW \citep{Koda:2016aa}.
Much deeper observations of a substantial spiral galaxy, even a single one,
provide an unbiased picture of molecular gas and cloud evolution
over a large galactic disk.

We present deep mosaic observations of the nearby 
barred spiral galaxy M83 (NGC 5236) in the fundamental CO(1-0) transition
with ALMA,
which provides a mass sensitivity and spatial resolution of $10^4~\Msun$ ($\gtrsim 3\sigma$) and $40~\pc$ respectively.
We use the ALMA 12m, 7m, and Total Power (TP) arrays,
convert the TP image into visibilities and
jointly-deconvolve them for high quality images \citep{Koda:2019aa}.
This paper presents the first results on the distribution of
molecular gas and clouds and their relation to star formation in M83.
Subsequent papers will present a more in-depth analysis on,
e.g., cloud properties and physical conditions as well as global and local environments of star formation.

This CO(1-0) study is complementary to large ALMA surveys of galaxies,
including PHANGS \citep{Sun:2018aa, Schinnerer:2019wp, Leroy:2021ab},
ALCHEMI \citep{Martin:2021aa, Harada:2019aa, Harada:2022aa},
VERTICO \citep{Brown:2021aa, Zabel:2022aa},
and ALMA JERRY (Jachym in private communication), as well as many other studies
of individual galaxies.

\subsection{The Barred Spiral Galaxy M83}\label{sec:introm83}

M83 is one of the closest barred spiral galaxies at the distance of $d=4.5\mpc$ \citep{Thim:2003aa}.
It closely resembles our own MW \citep{Churchwell:2009vn},
and its nearly face-on geometry, with an inclination angle of $i=26\deg$ (this study)
shows the full structure of the disk without the distance ambiguity that MW observations encounter.
Table \ref{tab:m83} summarizes the parameters of M83.
The galaxy's total stellar mass of $M_{\rm star}=2.5\times 10^{10}~\Msun$ 
and exponential disk scale length of $h=1.7\kpc$ \citep{Barnes:2014ut}, assuming a 
distance of 4.5 Mpc, are slightly smaller than those of the MW
\citep[$4.6\times 10^{10}~\Msun$ and $2.2\kpc$; ][]{Bovy:2013wd}.
The molecular gas mass of $M_{\rm H_2}=2.6\times 10^9~\Msun$ (this study) and
star formation rate of ${\rm SFR}=5.2\,~\Msun/\yr$ \citep{Jarrett:2019aa}
are slightly higher than those of the MW
\citep[$1.0\times 10^9~\Msun$ and $1.9\,~\Msun/\yr$; ][]{Heyer:2015qy, Chomiuk:2011wj}.
Therefore, M83 is only slightly smaller and slightly more active in star formation
than the MW.
The metallicity of M83 is 12+log(O/H)=$8.78\pm0.07$ in stars and $8.73\pm0.27$ in gas
at the galactocentric radius of $0.4 R_{25}$\citep{Bresolin:2016vw}.
It is also similar to the solar metallicity of 12+log(O/H)$_{\odot}$=8.66 \citep{Asplund:2005aa}.

This archetypal barred spiral galaxy has been a showcase
for multi-wavelength studies.
The Hubble Space Telescope (HST) imaged a large part of the optical disk with both wide and narrow band filters \citep{Blair:2014aa}.
With the proximity, these images permitted analyses of star clusters
\citep{Chandar:2010fk, Adamo:2015aa, Bialopetravicius:2020aa},
showing that high-mass clusters are less likely to form in the outer disk \citep{Adamo:2015aa},
and that clusters of all masses are disrupted in a timescale of a few 100~Myr in the disk \citep{Chandar:2010fk}.
The H$\alpha$ image was used to identify very young clusters with HII regions
\citep[$\lesssim 10\Myr$; ][]{Whitmore:2011lr}
and to study their interactions with the ISM \citep{Sofue:2018tu}.
The narrow and broad band HST images identify supernova remnants and their progenitors
\citep{Dopita:2010aa, Blair:2014aa, Williams:2019tj}.
The populations of supernova remnants and HII regions, as well as their feedback to the ISM,
were also studied in radio continuum emission \citep{Maddox:2006wv, Russell:2020tf}
and in X-ray \citep{Long:2014vl, Wang:2021aa}.
Three-dimensional optical spectroscopy covered large parts of the disk
\citep{Blasco-Herrera:2010wm, Poetrodjojo:2019aa, Hernandez:2021wv, Della-Bruna:2022ab, Della-Bruna:2022aa, Long:2022aa, Grasha:2022aa}.
These studies found a large contribution of diffuse ionized gas (DIG) to
the galaxy's H$\alpha$ flux \citep{Poetrodjojo:2019aa} and its radial and azimuthal variations:
the DIG fraction is lower in the center and spiral arms where the star formation is more active,
and higher in the interarm regions \citep{Della-Bruna:2022ab}.

Molecular gas in M83 has been observed in CO(1-0).
Early studies focused on particular portions of the galaxy,
such as the nucleus, bar, and spiral arms
\citep[e.g., ][]{Combes:1978tb, Wiklind:1990wy, Handa:1990ta, Lord:1991ts, Kenney:1991aa, Rand:1999vx}.
The whole disk was observed first with single-dish telescopes
\citep{Crosthwaite:2002yu, Lundgren:2004aa, Lundgren:2004pt},
which analyzed the CO $J$=2-1/1-0 line ratio between spiral arms
and interarm regions.
The most recent update with the ALMA single-dish telescopes concluded
that the ratio varies systematically from $\lesssim 0.7$
in low surface density interarm regions to $\gtrsim 0.7$
in higher surface density spiral arms
when it is analyzed at a 1~kpc-scale resolution \citep{Koda:2020aa}.

The emission from higher CO transitions and of other molecules has also been used
to trace the gas physical conditions mainly in the central region
\citep[e.g., ][]{Muraoka:2009aa, Tan:2018uf}.
These studies identified warm, dense molecular gas
in the central region \citep[e.g., ][]{Petitpas:1998wl, Sakamoto:2004aa, Muraoka:2009uy},
which could be due to gas collisions,
that accelerated the chemical evolution \citep{Harada:2019aa, Martin:2009aa}.
M83 was also used as a test bed of [CI] line transitions
as a potential alternative tracer of molecular regions and their masses
with some success \citep{Jiao:2021aa} and with caution \citep{Miyamoto:2021wl}.

More recent CO(1-0) observations resolve molecular structures
on sub-kpc scales \citep[$\sim 100\pc$; ][]{Hirota:2014wd} and
even on a cloud scale ($\sim 40\pc$) in a part of the disk
\citep{Hirota:2018aa}.
A portion of the eastern spiral arm, the part stretching
from the bar-end toward east, was resolved into two parallel molecular ridges.
Molecular clouds along one of the ridges show systematically higher star formation efficiency
than those on the other ridge \citep{Hirota:2018aa},
suggesting that the star formation trigger is related to the large-scale pattern.
\citet{Egusa:2018aa} found that the gas velocity dispersion is enhanced in the bar,
compared to the disk, potentially spreading the gas and suppressing star formation there.
The majority of clouds without HII regions are in the interarm regions \citep{Hirota:2018aa}.
These results, obtained in a small part of the disk, suggest that
large-scale galactic structure controls the properties of molecular clouds and star formation.
The masses of star clusters appear to be determined by those of their parental clouds.
Indeed, the mass functions of molecular clouds and clusters appear to track each other
in their upper mass cutoffs \citep{Freeman:2017aa} or in their slopes \citep{Mok:2020aa}.

The CO(1-0) data presented in this study show both large-scale galactic structures as well as molecular clouds.
It enables an unbiased census of cold molecular gas,
and mapping the cloud mass spectrum as a function of galactic structure.
The fully-resolved CO(1-0) map can potentially reveal conditions
in the immediate vicinity of star forming regions.
The resolution of $\sim 40\pc$ is crucial for these studies,
as it is the scale of a typical molecular cloud (diameter of 40~pc),
not the scale of a typical separation between clouds
\citep[$\gtrsim 100\pc$; ][]{Sun:2018aa, Sun:2020aa, Sun:2020tt, Rosolowsky:2021aa}.

\begin{deluxetable}{lcc}
\tablecaption{Global Properties of M83 \label{tab:m83}}
\tablehead{
\colhead{Parameter} & \colhead{Value} & \colhead{Ref.}}
\startdata
NGC                     & 5236  &      \\
Morphology              & SAB(s)c &  1     \\
R.A.(J2000)             & $13^\mathrm{h} 37^\mathrm{m} 0\fs 57$  &  2    \\
DEC(J2000)              & $-29\arcdeg 51\arcmin 56\farcs 9$ &  2     \\
Distance (Mpc)          & 4.5  &    3  \\
m-M (mag)               & 28.25$\pm$0.15  &    3    \\
$V_{\rm sys}$ ($\kmps$) & $511\pm 3$  & 4      \\
$V_{\rm rot}$ ($\kmps$) & $174\pm 10$  & 4      \\
$\Omega_{\rm b}$ ($\kmps/\kpc$)    & $57.4\pm 2.8$   & 5 \\
$R_{\rm CR}$ ($\kpc$)   & 3.0   & 4,5 \\
$M_{\rm B}$ (mag)       & -20.14  &  6,3   \\
B-V (mag)               & 0.59  & 6 \\
$D_{\rm 25}$ (arcmin)   & 12.9  &  1     \\
$D_{\rm 25}$ (kpc)      & 16.9  &  1,3   \\
Axis Ratio              & 1.12  &  1 \\
P.A. (deg)              & $225 \pm 1$  & 4    \\
Inclination (deg)       & $26 \pm 2$   & 4       \\
$M_{\rm star}$ ($\Msun$)& $2.5\times 10^{10}$  &  7   \\
$\Mmol$ ($\Msun$) & $2.6\times 10^9$ &  4     \\
$M_{\rm HI}$ ($\Msun$)  & $7.8 \times 10^9$  & 8   \\
SFR ($\Msun\,{\rm yr}^{-1}$)  & 5.17$\pm$1.79  &  7      \\
12+log(O/H)$_{\rm star}$             & 8.78$\pm$0.07  & 9 \\
12+log(O/H)$_{\rm gas}$             & 8.73$\pm$0.27  & 9 \\
$F_{\rm TIR}$ ($10^{-12}\,\rm erg\,s^{-1}cm^{-2}$) & 4.07$\pm$0.36 & 10
\enddata
\tablecomments{The galaxy center is obscured and is difficult to determine.
We adopt the center coordinate
derived as the symmetry center of $K$-band isophotes
within the central arcmin region \citep{Thatte:2000aa, Diaz:2006aa, Sakamoto:2004aa}.
The stellar and gas metallicities are the values at a galactocentric radius of $0.4 R_{25}$ ($=0.2 D_{25}$).
All the values are corrected for the adopted distance when necessary.
Reference:
(1) \citealt{de-Vaucouleurs:1991lr},
(2) \citealt{Thatte:2000aa},
(3) \citealt{Thim:2003aa},
(4) This study,
(5) \citealt{Hirota:2014wd},
(6) \citealt{Cook:2014aa},
(7) \citealt{Jarrett:2019aa},
(8) \citealt{Koribalski:2018vj},
(9) \citealt{Bresolin:2016vw},
(10) \citealt{Marble:2010aa}.
}
\end{deluxetable}

\section{Observations}\label{sec:obs}

The nearby barred spiral galaxy M83 was observed
with the 12m, 7m, and Total Power (TP) arrays of ALMA in CO(1-0)
under project code 2017.1.00079.S.
\footnote{The 7m and TP arrays are also called the Atacama Compact Array (ACA),
or the Morita array named after Professor Koh-ichiro Morita,
a member of the Japanese ALMA team and designer of the ACA.}

\begin{figure}[h]
\epsscale{1.1}
\plotone{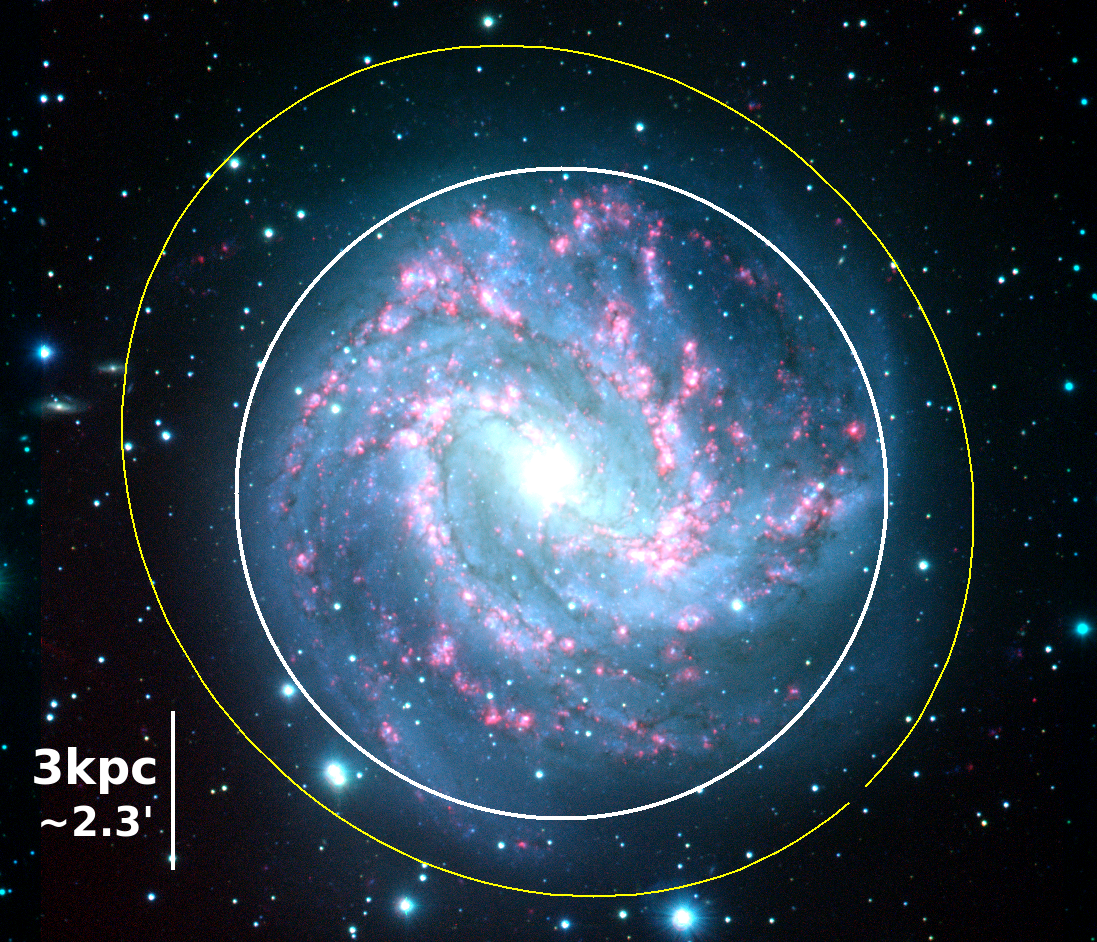}
\caption{
An optical B-V-H$\alpha$ composite image of M83 taken at the Cerro Tololo Inter-American Observatory (CTIO)
and downloaded from the NASA/IPAC Extragalactic Database (NED).
The white circle is the region observed with the ALMA 12m and 7m arrays and encloses
the whole star-forming H$\alpha$ disk of M83.
The ALMA Total Power array observed an area larger than the white circle (see Figure \ref{fig:pointings}).
The yellow ellipse shows the optical edge of the disk at $R_{25}$ \citep{de-Vaucouleurs:1991lr}.
\label{fig:optimage}}
\end{figure}

\subsection{12m and 7m Arrays}

The whole star-forming disk of M83 was mapped
with a 435-pointing mosaic using the 12m and 7m arrays (Figures \ref{fig:optimage} and \ref{fig:pointings}).
Although the ALMA's default mosaic pattern uses
different sets of pointing positions for the two arrays to account for the different primary beam sizes,
we used the same pointing positions for the two arrays.
In this way, each pointing position could be imaged with the 12m and 7m array data
separately and together, which is useful for consistency checks.

This mosaic setup observes more positions with the 7m array than the ALMA's default
and over-samples the area with its primary beam.
However, it requires roughly the same amount of observing time with no loss in sensitivity across the mosaic area,
as the array spends a shorter amount of time at each pointing.
The overhead associated with the over-sampling is almost entirely from antenna slew,
which is short and negligible with ALMA.
Hence, the benefit outweighs the drawback.
We successfully mapped M83 in this custom manner (with considerable support
from the North American ALMA Science Center [NAASC] staff for this setup).

The total of 435 pointing positions were split into six sub-regions
(three columns and three rows) and into 6 scheduling blocks (SBs),
given the observatory constraint of a maximum of 150 pointings per SB.
Figure \ref{fig:pointings} shows the design of the 6 SBs.
The column and row arrangements overlap each other
to check consistency among SBs.
For convenience, we name the SBs
\textit{column 0-2} from left to right (east to west) and \textit{row 0-2} from top to bottom (north to south).
Each of the SBs was observed multiple times, each of which is called an execution block (EB).
20 (78) EBs were executed on the 12m (7m) array,
each of which passed the observatory quality assurance process;
the breakdown is shown in  Table \ref{tab:EBs}.
Each EB included observations of bandpass, gain, and flux calibrators
as well as the target galaxy.

\begin{deluxetable}{cccc}
\tablecaption{Number of Execution Blocks (EBs) \label{tab:EBs}}
\tablehead{
\multicolumn{2}{c}{Scheduling Block (SB)} & \colhead{12m} & \colhead{7m}}
\startdata
  column & 0 & 5 & 13   \\
         & 1 & 3 & 13   \\
         & 2 & 3 & 13  \\
  row    & 0 & 3 & 13   \\
         & 1 & 3 & 13   \\
         & 2 & 3 & 13          
\enddata
\end{deluxetable}

The typical system temperatures of EBs range from 96 to 116~K, with an average of 105~K for the 12m array,
and from 91 to 138~K, with an average of 101~K, for the 7m array.
The setup of correlators (i.e., spectral windows) is listed in Table \ref{tab:obscorr}.
The velocity coverage and channel width are enough to cover the full velocity
width of M83 at a 0.3$\kmps$ resolution.

\begin{deluxetable*}{ccccccccc}
\tablecaption{ALMA Spectral Window Setups \label{tab:obscorr}}
\tablehead{
\colhead{} & \colhead{} & \multicolumn{3}{c}{Frequency} & \colhead{}& \multicolumn{3}{c}{Velocity}  \\
\cline{3-5} \cline{7-9}
\colhead{Array} &  \colhead{$N_{\rm chan}$} & \colhead{Band Width} & \colhead{Chan. Increment} & \colhead{Resolution} & \colhead{} & \colhead{Band Width} & \colhead{Chan. Increment} & \colhead{Resolution}  \\
\colhead{} & \colhead{} & \colhead{(MHz)} & \colhead{(kHz)} & \colhead{(kHz)} & \colhead{} & \colhead{($\kmps$)} & \colhead{($\kmps$)} & \colhead{($\kmps$)} }
\startdata
  12m & 3840 &  234.375 & 61.035 & 122.070 && 609.559 & 0.15874 & 0.31748\\
   7m & 4096 &  250.000 & 61.035 & 122.070 && 650.196 & 0.15874 & 0.31748\\
   TP & 4096 &  250.000 & 61.035 & 122.070 && 650.196 & 0.15874 & 0.31748\\
\enddata
\end{deluxetable*}

\subsection{Total Power Array}

\begin{figure}[h]
\epsscale{1.1}
\plotone{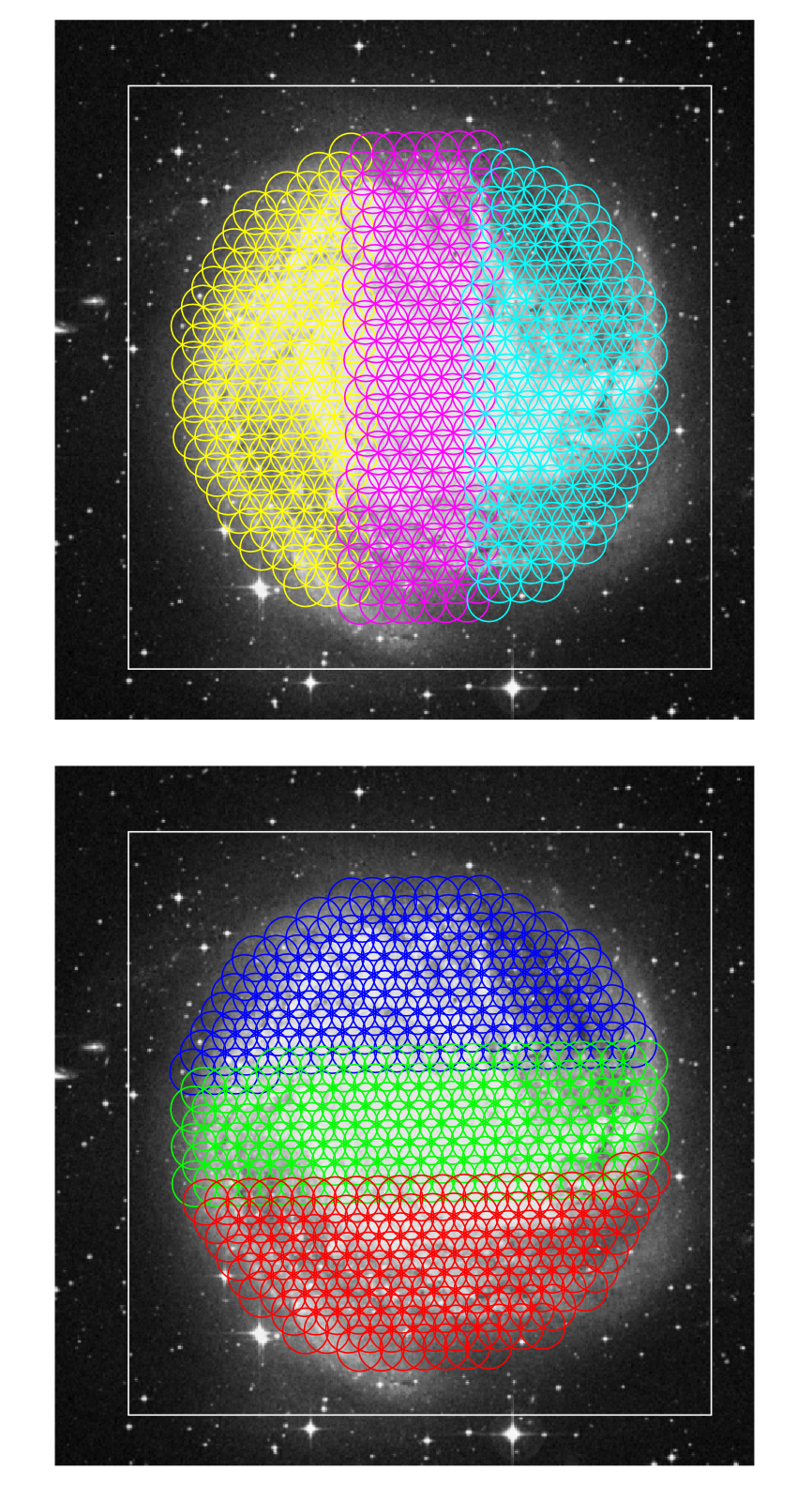}
\caption{
The 435-pointing mosaic pattern of the ALMA 12m and 7m arrays.
The optical image in the background is from the Digitized Sky Survey.
Each circle represents the primary beam size of the 12m array.
The pointings are split into 6 redundant science blocks:
three columns (top panel) and three rows (bottom panel).
The white box indicates
the $11.7\arcmin \times 11.7\arcmin$ area
mapped by the Total Power (TP) array.
The 12m and 7m arrays roughly cover a circular region
with a diameter of $9.4\arcmin$ at an almost constant
sensitivity.
\label{fig:pointings}}
\end{figure}

The TP observations were carried out using the On-The-Fly (OTF) mapping technique \citep{Mangum:2007aa, Sawada:2008xz}.
Two SBs were configured to scan the whole galaxy at once either in the RA or DEC directions (the white box in Figure \ref{fig:pointings}),
instead of the standard ALMA setup of arranging six separate SBs corresponding to those of the 12m and 7m arrays.
This minimizes scan errors when the SBs are combined.
An OFF position was integrated for 6.336 sec every OTF scan (20.160 sec integration per raster scan).
Each SB covered a $11.7\arcmin \times 11.7\arcmin$ area,
which extends beyond the molecular disk of M83 (the white box in Figure \ref{fig:pointings}).
The data contain the emission-free sky on all edges.
The ACA correlator was used and the parameters are listed in Table \ref{tab:obscorr}.

The SBs were executed repeatedly with multiple EBs.
The total of 125 EBs obtained the Quality Assurance stage 0 (QA0) status of Pass or Semi Pass according to the ALMA observatory records.
The amplitude calibration was performed with the standard chopper-wheel method.
We rejected 5 EBs due to bad weather (unreasonably low intensity),
spurious pointing corrections (blurred map appearance),
and relatively large flux errors with respect to the other EBs (deviations greater than a few \%).
We used 120 EBs after the rejections.
The average number of TP antennas participating in the observations was 3.71 out of 4.
The total observing and on-source times were 110.1 h and 60.4 h, respectively.
The system temperature
$T_{\rm sys}$ was 99, 105, 113 (median), 121, and 129 K at 10th, 25th, 50th, 75th, and 90th percentiles.

\section{Data Reduction}\label{sec:reduction}

We used the Common Astronomy Software Application \citep[CASA: ][]{McMullin:2007aa, Bean:2022aa} version 5.1
for calibration (this section), and the Multichannel Image Reconstruction, Image Analysis,
and Display \citep[MIRIAD; ][]{Sault:1995kl, Sault:1996uq} for imaging (Section \ref{sec:imaging}).

\subsection{12m \& 7m-Arrays}\label{sec:red12m}

The observatory performed quality assurance on each dataset and delivered a calibration pipeline script with each data delivery.
We ran them and checked the calibrated amplitudes, phases, and fluxes of bandpass and gain calibrators.\footnote{The ALMA observatory announced an error in amplitude calibration in the 12m and 7m array observations, which is introduced during the observations due to a poor calibration strategy. We neglect it because its impact is small, and is, at most, only about 0.7\% at the galactic center in the velocity channels where the emission is the strongest.
This error is calculated as the ratio of the highest antenna temperature $T_{a}^*$ in the TP cube (Section \ref{sec:redtp}) and typical system temperature $T_{\rm sys}$.}
We did not subtract continuum emission, but checked that it was not detected in the final data cubes.

\subsection{Total Power Array}\label{sec:redtp}

Individual spectra were calibrated by mostly following the CASA calibration pipeline
except for the spectral baseline subtraction.
We subtracted the baselines later, from data cubes rather than from individual spectra.

The calibrated spectra were re-sampled on a grid with pixel size of $5.62\arcsec$
using the prolate spheroidal function with a size of 6 pixels \citep{Schwab:1980aa, Schwab:1984}.
The effective full-width half maximum (FWHM) beam size after this regridding -- hence smoothing -- is $56.6\arcsec$.
We generated separate data cubes for the 120 EBs, calculated the flux ratios and errors
of all of their pairings, and solved for relative flux scales by inverting the design matrix.
The derived 120 scaling coefficients, whose median was set to 1, have a small scatter of only 1.3\%.
We applied these coefficients to co-add all the spectra into two data cubes
of the RA and DEC scans.
The absolute flux scale was monitored repeatedly by the ALMA observatory,
and the consistency in time and frequency was checked
not only within our frequency band (Band 3), but also against other bands.

Spectral baselines were subtracted with a straight line fit in these cubes.
The data reduction pipeline uses a high-order polynomial fit to individual spectra,
which sometimes causes artifacts.
In practice, the baselines are mostly flat at the observed frequency.
A straight-line fit is sufficient.
It can be applied to individual spectra as done in the pipeline, or equivalently, to data cubes
after their integration, as we did in this study.
The latter is computationally much more efficient.

\begin{figure}[h]
\epsscale{0.8}
\plotone{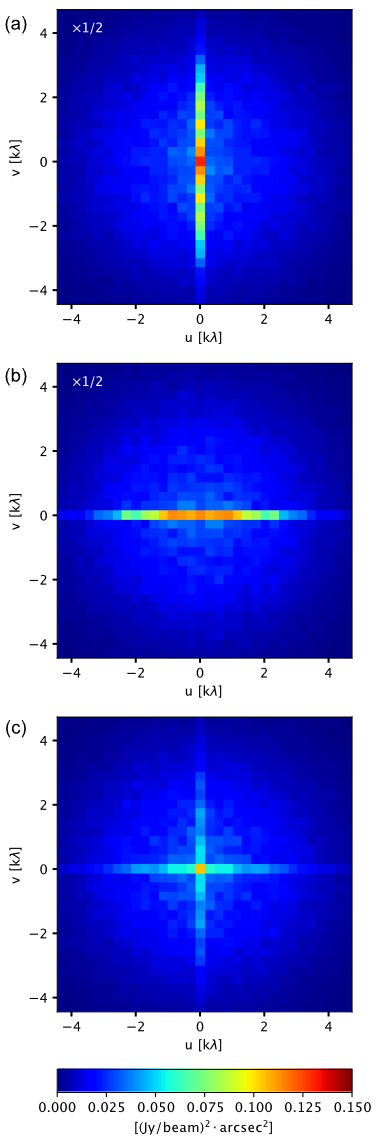}
\caption{Power spectrum density maps of TP observations for
(a) the RA scan, (b) DEC scan, and (c) combination of the RA and DEC scans.
The power spectra of 200 emission-free channels with a channel width of 61.6~kHz are averaged.
Panels (a) and (b) include twice less integration than panel (c),
and hence, their noise power densities are scaled by 1/2
so that their color scales can be compared directly to that of panel (c).
\label{fig:tppowerspec}}
\end{figure}

The RA and DEC data cubes were combined with the \citet{Emerson:1988hz} method.
Figure \ref{fig:tppowerspec} compares the noise power spectrum densities
of the individual RA and DEC scans and of their combination.
Systematic noise in the scan directions very clearly appear as
excess noise power in the vertical and horizontal directions in panels (a) and (b), respectively.
This noise is mainly due to the shared OFF integration in each OTF scan \citep[see ][]{Heyer:1998lr, Jackson:2006ya}.
The combination of the RA and DEC scans reduce the noise power by a factor of about 2.

The antenna temperature $T_{a}^*$ was converted to the Jy/beam unit
by multiplying the coefficient, $C_{Jy/K} = 40.7$ Jy/K, measured by the observatory.
This coefficient corresponds to a main beam efficiency of $\eta_{\rm mb}=0.856$
via the equation
\begin{equation}
    \eta_{\rm mb} = \frac{2 k_{\rm b} \Omega_{\rm TP}}{\lambda^2} \frac{1}{C_{Jy/K}},
\end{equation}
where $k_{\rm b}$ is the Boltzmann constant, $\Omega_{\rm TP}$ is the effective beam area,
and $\lambda$ is the observing wavelength.
We used a beam size of FWHM = $56.6\arcsec$ ($\approx 2.74\times 10^{-4} \,\rm rad$)
at $\lambda=2.60\,\rm mm$, and
\begin{equation}
 \Omega_{\rm TP}=\frac{\pi \rm FWHM^2}{4\ln 2}.
\end{equation}

The final cube has a root-mean-square noise of $6.2$ mK in $T_{\rm mb}$,
and 0.25~Jy/beam, in a velocity channel width of $0.159\kmps$.
The total integrated flux in the cube is
$S_{\nu} dv =1.67\times 10^4 \Jykmps$.
This map was used to show the systematic variations of CO $J=$2-1/1-0 ratio
over the disk of M83 \citep{Koda:2020aa}.

\subsection{Total Power Cube to Visibilities}\label{sec:tp2vis}

The TP data cube was converted to the form of interferometric data (visibilities)
using the Total Power to Visibilities package \citep[TP2VIS; ][]{Koda:2019aa}.
This conversion permits a joint imaging (deconvolution) of the 12m, 7m, and TP
data with existing imaging algorithms and software for interferometers.
By construction, the imaging algorithms work much better when starting with
a higher quality dirty image (i.e., more complete \textit{uv} coverage).
The combination of 12m, 7m, and TP data before imaging ensures this higher quality,
and motivates this approach.

Two modifications were made to the method presented in \citet{Koda:2019aa}.
First, when we deconvolved the TP cube with the TP primary beam,
we also took into account the smoothing kernel introduced in re-gridding
the observed data onto the data cube grid
\citep[the first step in TP2VIS, i.e., step A in ][]{Koda:2019aa}.
That is, the TP primary beam that we used for the deconvolution
is a convolution of a Gaussian beam and the prolate spheroidal function (\S \ref{sec:redtp}).

\begin{figure}[h]
\epsscale{1.0}
\plotone{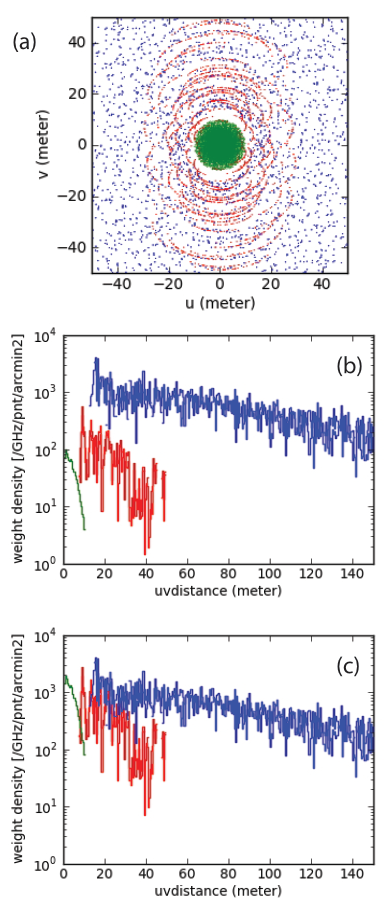}
\caption{
Relative weight densities of the 12m, 7m, and TP data in the \textit{uv} space
for a pointing at the center of mosaic:
(a) the \textit{uv} distributions of 12m (blue), 7m (red), and TP (green) data,
(b) their weight distributions from the observations, and
(c) after the scaling.
\label{fig:relweights}}
\end{figure}

The relative weight densities among the 12m, 7m, and TP data from the observations
are not matched as a function of uv-distance (Figure \ref{fig:relweights}b).
Their relative observing times are fixed by the observatory,
but it causes a shortage in sensitivity for the 7m and TP data
with respect to that of the 12m data.
Thus, the weights of the 7m and TP visibilities are scaled up
by factors of 5 and 20, respectively.
The factors are determined by the inspection of Figure \ref{fig:relweights}b.
\footnote{This is equivalent to down-weighting high quality data,
that is, scaling down the weights of the 12m and 7m visibilities
by factors of 1/20 and 1/4 (=5/20), respectively, to match those of the TP data.}
Figure \ref{fig:relweights}a shows the central part of \textit{uv} coverage
by the 12m (blue), 7m (red), and TP visibilities (green).
Figure \ref{fig:relweights}b shows their weight distributions
as a function of \textit{uv} distance.
The optimal relative weight distribution would show smooth transitions
from the 12m to 7m, then to TP data, which would ensure 
a more optimal PSF for imaging.
Figure \ref{fig:relweights}c shows the distributions after the weight scaling.

The manipulation of the imaging weights as described in the previous paragraph
may appear somewhat arbitrary.
We note that the imaging of radio interferometer data
routinely involves manipulating weights. For instance, employing robust or uniform weighting
during imaging enables a user to tune the spatial resolution and sensitivity
according to the science goals of the project.
The adjustment in this study follows this precedent.
Further, the \textsc{Feather} task in CASA is widely used to
combine TP image with 12m and/or 7m image
by \textit{implicitly} adopting a weight scaling similar to ours.
\textsc{Feather} adds the two images without accounting for
their sensitivity difference.
It adjusts the relative weights so that their combined weight densities
reproduce the ones defined by the convolution beam of
the CLEANed 12m and/or 7m image (i.e., the restoring beam in CASA's \textsc{tclean};
note that the Fourier transformation of this beam is the weight distribution in the $uv$ space).
This is equivalent to scaling up/down the weight of the TP image.
The \textsc{sdintimaging} task in CASA also includes this implicit scaling.

\section{Imaging}\label{sec:imaging}

The 12m, 7m, and TP array data were jointly-deconvolved with
the Steer CLEAN algorithm (the \textsc{invert} and \textsc{mossdi2} tasks)
in the Multichannel Image Reconstruction, Image Analysis, and Display (MIRIAD) software
package \citep[][this choice of software is explained in Section \ref{sec:miriad}]{Sault:1995kl, Sault:1996uq}.
The visibility data from CASA were converted to the MIRIAD data format.
For simplicity, we adopted a single average $T_{\rm sys}$ value for each EB,
as it does not vary much within an EB in ALMA Band 3 observations.
$T_{\rm sys}$ was used for weighting in imaging.
We applied the factors of 5 and 20 to scale up the weights of the 7m and TP data
as discussed in Section \ref{sec:tp2vis}.
The whole 435-point mosaic field was CLEANed at once.
We made three data cubes with three channel widths: 5, 2, and 1$\kmps$.
We use the $5\kmps$ cube and its parameters in the rest of the paper
unless otherwise stated.
The parameters of the three data cubes are listed in Table \ref{tab:cubes}.

The Briggs ``robust" weight-control scheme was also employed in addition to the $T_{\rm sys}$ weighting.
We measured the sensitivity and beam size at various robust values and
adopted ${\rm robust}=-0.2$ as the optimal value for this study.
The cell size is set to $0.25\arcsec$.
The root-mean-square (RMS) noise is 3.8 mJy (96~mK) in a $5\kmps$ channel.
The point spread function (PSF) size is $2.12\arcsec \times 1.71\arcsec$
at a position angle of $-89.0\arcdeg$.

We ran CLEAN in two steps:
(1) first ran it down to a $1\sigma$ flux level without a mask, and
(2) then made a mask for large areas of contiguous pixels at $>0.8\sigma$ significance
and ran CLEAN again on the residual map from step (1).
The total flux of the residual map naturally decreased with the CLEAN iteration.
We stopped it when the average intensity within the mask hit the floor or $\sim 0.2\sigma$.
The last step was implemented because extended emission, even at a very low level,
could amount to a large total flux over large areas.
The residual map from step (1) contains about 6.6\%
of the total flux of the dirty map before CLEAN, and that from step (2) contains about 1.5\%.

A difference in beam area between 
the synthesized PSF from the \textit{uv} coverage 
and the gaussian PSF that is applied to clean components
causes a problem in flux measurement \citep{Koda:2019aa}.
To mitigate this problem, we matched the PSF areas with the scheme presented by \citet{Jorsater:1995aa}.
In effect, the residual map from step (2), hence the 1.5\% of the total flux,
was scaled (multiplied) by a factor based on the ratio of beam areas (0.96, Table \ref{tab:cubes}) and added to the clean component map.
This step is necessary to ensure the flux accuracy in the final data cube.
The total integrated flux in the cube is $S_{\nu} dv =1.68\times 10^4 \Jykmps$, which is consistent with that of the TP data cube with a 0.6\% error (Section \ref{sec:redtp}).
We refer to this final cube as ``the cube", ``the 12m+7m+TP cube", or ``the combined cube".

\begin{deluxetable*}{cccccccccccc}
\tablecaption{Parameters of Data Cubes \label{tab:cubes}}
\tablehead{
\colhead{(1)} & \colhead{(2)} & \colhead{(3)}& \colhead{} & \colhead{(4)} & \colhead{} & \multicolumn{2}{c}{(5)} & \colhead{} & \colhead{(6)} & \colhead{(7)} & \colhead{(8)}\\
\colhead{} & \colhead{} & \colhead{Velocity}& \colhead{} & \colhead{Beam Size} & \colhead{} & \multicolumn{2}{c}{Noise ($1\sigma$)} & \colhead{} & \colhead{Residual/Total} & \colhead{Beam Ratio} & \colhead{Flux Recovery} \\
\cline{3-3} \cline{5-5} \cline{7-8}
\colhead{Name} & \colhead{$N_{\rm chan}$} & \colhead{Start, End, CW} & \colhead{} & \colhead{$b_{\rm maj}$, $b_{\rm min}$, P.A.} & \colhead{} & \colhead{$\delta S_{\nu}$} & \colhead{$\delta \Tb$} & 
\colhead{} & \colhead{} & \colhead{} & \colhead{} \\
\colhead{} & \colhead{} & \colhead{($\kmps$)}  & \colhead{} & \colhead{($\arcsec$, $\arcsec$, $\arcdeg$)} & \colhead{} & \colhead{(mJy/beam)} & \colhead{(mK)} & \colhead{} & \colhead{(\%)} & \colhead{} & \colhead{}}
\startdata
  CUBE5 & 100 & 250, 745, 5 && 2.12, 1.71, -89.0 && 3.8 &  96 &&  1.5 & 0.96 & 1.01 \\
  CUBE2 & 250 & 250, 748, 2 && 2.12, 1.71, -89.0 && 5.8 & 148 &&  2.5 & 0.96 & 1.00 \\
  CUBE1 & 310 & 340, 649, 1 && 2.09, 1.68, +89.9 && 7.7 & 195 &&  2.7 & 0.96 & 1.00 \\
\enddata
\tablecomments{The Briggs's robust parameter is set to $-0.2$ for all cubes. $\Tb {\rm\, [K]} =25.4S_{\nu} {\rm\, [Jy]}$.
(1) Name of cube.
(2) Number of channels.
(3) Start and end velocities, and channel width (CW).
(4) Major and minor axis diameters and position angle of the beam.
(5) Root-mean-squre (RMS) noise in a channel in units of Jy/beam and K.
(6) Fraction of flux left in residual map after the two-step CLEAN procedure. The total flux is calculated in the dirty cube before CLEAN.
(7) Areal ratio of convolution and synthesized beams.
(8) Ratio of total fluxes in 12m+7m+TP and TP cubes. The beam area correction is applied to the residual part of the 12m+7m+TP cube.
CUBE1 shows a slightly different beam size from the other two due to the different velocity coverage.
}
\end{deluxetable*}

\subsection{Conversion from Jy/beam to K}

The data cube was converted from
intensity (brightness) $I_{\nu}$, in units of Jy/beam, to
Rayleigh-Jeans brightness temperature $\Tb$ in units of K.
From the Rayleigh-Jeans equation of
$I_{\nu} = 2 k_{\rm B} T /\lambda^2$
and the relation between the intensity and
flux density $S_{\nu}$ of $I_{\nu} = S_{\nu}/ \Omega_{\rm beam}$,
the conversion equation is
\begin{eqnarray}
    \Tb &=& \frac{\lambda^2}{2 k_{\rm B} \Omega_{\rm beam}} S_{\nu} \\
      &=& 13.6 \,{\rm K} \left( \frac{\lambda}{\rm 1mm} \right)^2 \left( \frac{b_{\rm maj} \times b_{\rm min}}{\rm 1\arcsec \times 1\arcsec} \right)^{-1} \left( \frac{S_{\nu}}{\rm 1 Jy} \right), \label{eq:convStoT}
\end{eqnarray}
where $k_{\rm B}$ is the Boltzmann constant,
$\Omega_{\rm beam}$ is the beam area in steradians,
and $b_{\rm maj}$ and $b_{\rm min}$ are the FWHM sizes of the beam
along its major and minor axes in arcsec.
The $S_{\nu}$ is a flux density \textit{within} a beam and
can be regarded as equivalent to the $I_{\nu}$ in the data cube.

In our observations we have
$\lambda=2.60\rm\, mm$, $b_{\rm maj}=2.12\arcsec$, and $b_{\rm min}=1.71\arcsec$, and hence, $\Tb =25.4S_{\nu}$.

\subsection{MIRIAD Over CASA for Mosaic Imaging} \label{sec:miriad}

We used MIRIAD for imaging.
MIRIAD works better than CASA primarily because CASA's \textsc{tclean} task,
as of version 6.4, does not account for spatial variations of PSF across the mosaic,
even though the \textit{uv} coverage varies among the mosaic pointing positions.
Hence, CASA's  \textsc{tclean} uses an incorrect PSF for deconvolution in mosaic observations.
Because of this, we often ran into runaway divergences in \textsc{tclean}, most likely,
but not always, at
the edges of the mosaic
which is covered either by the 12m or 7m array primary beam, but not both.
Some ``tricks", such as \textsc{automasking} or a smaller \textsc{cycleniter} value,
often obscure the problem,
but it is unclear if they maintain accuracy since the underlying PSF is still incorrect.
Even the new task \textsc{sdintimaging} \citep{Rau:2019aa}
does not account for the spatially-variant PSF.

At this moment, we regard MIRIAD's imaging tasks as more reliable than CASA's
at least for a medium size or larger mosaic.
The MIRIAD imaging tasks have been used successfully in joint-deconvolution
of mosaic data from heterogeneous array interferometers and single dish telescopes
\citep[e.g., ][]{Koda:2009wd, Koda:2011nx, Momose:2010cr, Momose:2013aa, Donovan-Meyer:2012fk, Donovan-Meyer:2013uq, Hirota:2018aa, Kong:2018aa, Sawada:2018aa}.
The problem of CASA may be less notable, or even negligible, for smaller mosaics,
where all pointing positions are observed in a sequence within a short cycle,
since the \textit{uv} coverages are similar among the positions in such observations.

\section{Maps} \label{sec:maps}

The reduced data is a 3-D cube with axes of right ascension, declination, and velocity.
To make 2D moment maps,
we generate two mask cubes from the data cube to include compact and extended emission components,
and combined them into a single mask cube.
The first mask is defined with the regions above $>2\sigma$ including
at least one $>4\sigma$ peak in 3-D.
The regions are expanded spatially outward from their edge by one PSF major-axis diameter to include the envelopes of the emission.
This mask is designed to include compact components detected in the original data.
The second mask is made in the same way, but with the data cube smoothed to a $5\arcsec$ resolution to recover 
diffuse, extended emission
around more significant emission.
The combined mask cube identifies only a few \% of the pixels in the cube as emission pixels and the rest as noise. Hence, without the mask, the moment maps suffer significantly from the noise.

\begin{figure*}[h]
\epsscale{1.2}
\plotone{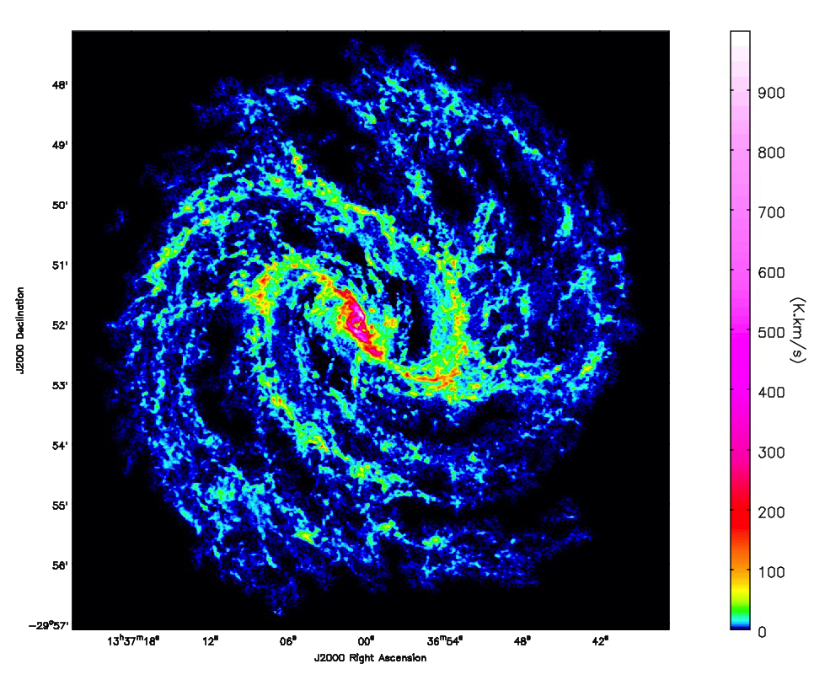}
\caption{
The CO(1-0) integrated intensity $I_{\rm CO}$ map (the ``mom0" map).
At a distance of 4.5~Mpc, an angle of $1\arcmin$ corresponds to 1.3~kpc, and that of $1\arcsec$ corresponds to 21.8~pc.
The ALMA 12m and 7m-array observations roughly cover a $9.4\arcmin$-diameter (12.3~kpc) circular region (see Figure \ref{fig:pointings}).
The beam size is $2.12\arcsec \times 1.71\arcsec$ ($46.3 \pc \times 37.3 \pc$) at a position angle of $-89.0\arcdeg$.
The noise level varies spatially, as a mask is used to generate this map
(as a guideline, a pixel with only one channel integrated has an RMS noise of $0.48\,\rm K\kmps$).
\label{fig:combmom0}}
\end{figure*}

\begin{figure*}[h]
\epsscale{0.85}
\plotone{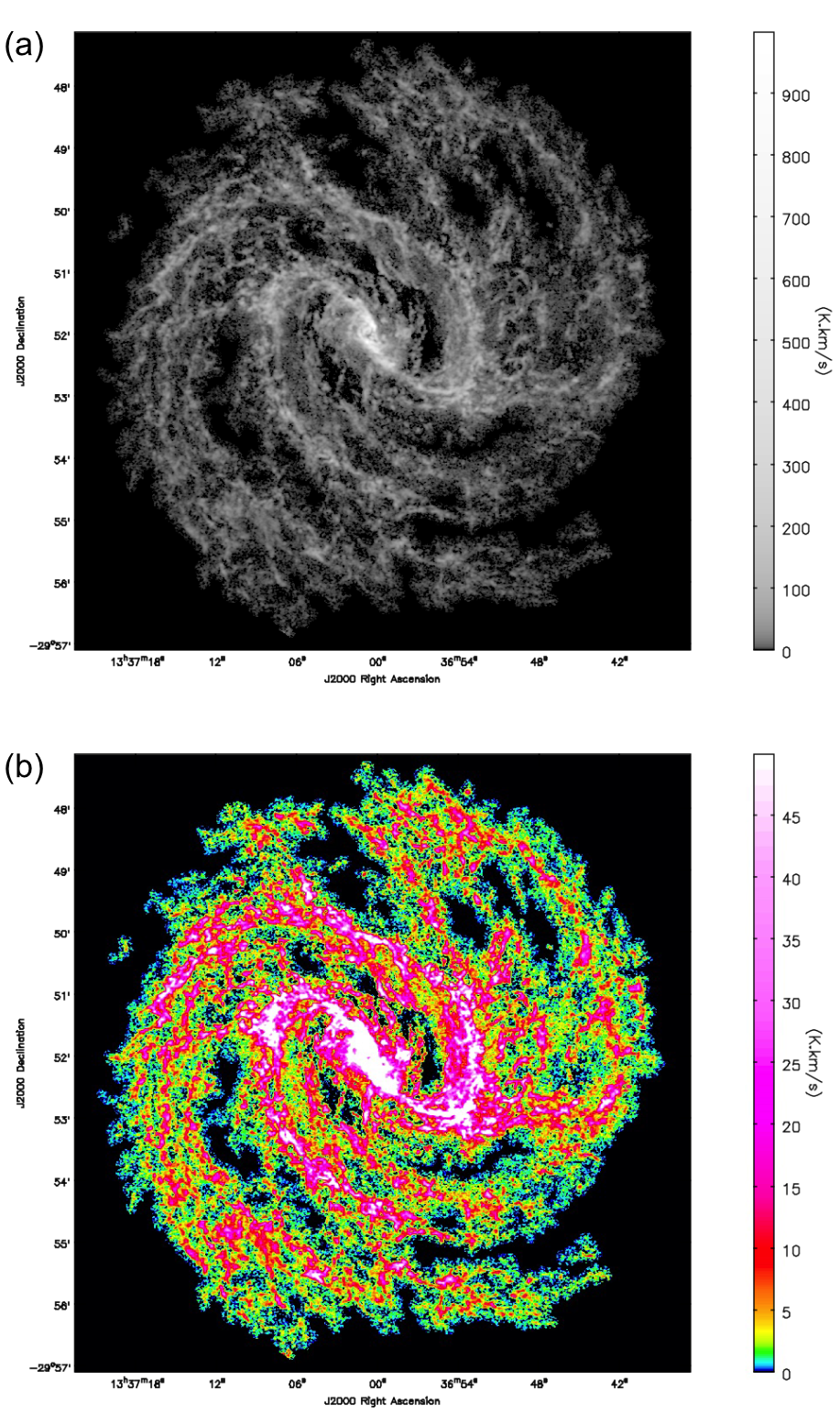}
\caption{
The same as Figure \ref{fig:combmom0}, but (a) in grayscale, and (b) with a different range of integrated intensity.  These images show faint structures clearer than Figure \ref{fig:combmom0}. Panel (b) shows the regions where the integrated intensity is above and below the average within typical molecular clouds ($I_{\rm CO}\approx 45 \,\rm K \kmps$).
\label{fig:combmom0alternative}
}
\end{figure*}

Figures \ref{fig:combmom0} and \ref{fig:combmom0alternative}
show the CO(1-0) integrated intensity $I_{\rm CO}$ maps (also called the mom0 map)
in color and gray in the full intensity range, and in a limited intensity range, respectively.
These three versions of the $I_{\rm CO}$ map are presented to show diverse structures at different intensities.
Figure \ref{fig:combmom0alternative}b differentiates the intensity above and below an intensity of typical clouds in the Milky Way ($\sim 50 \,\rm K \kmps$).

\begin{figure*}[h]
\epsscale{0.85}
\plotone{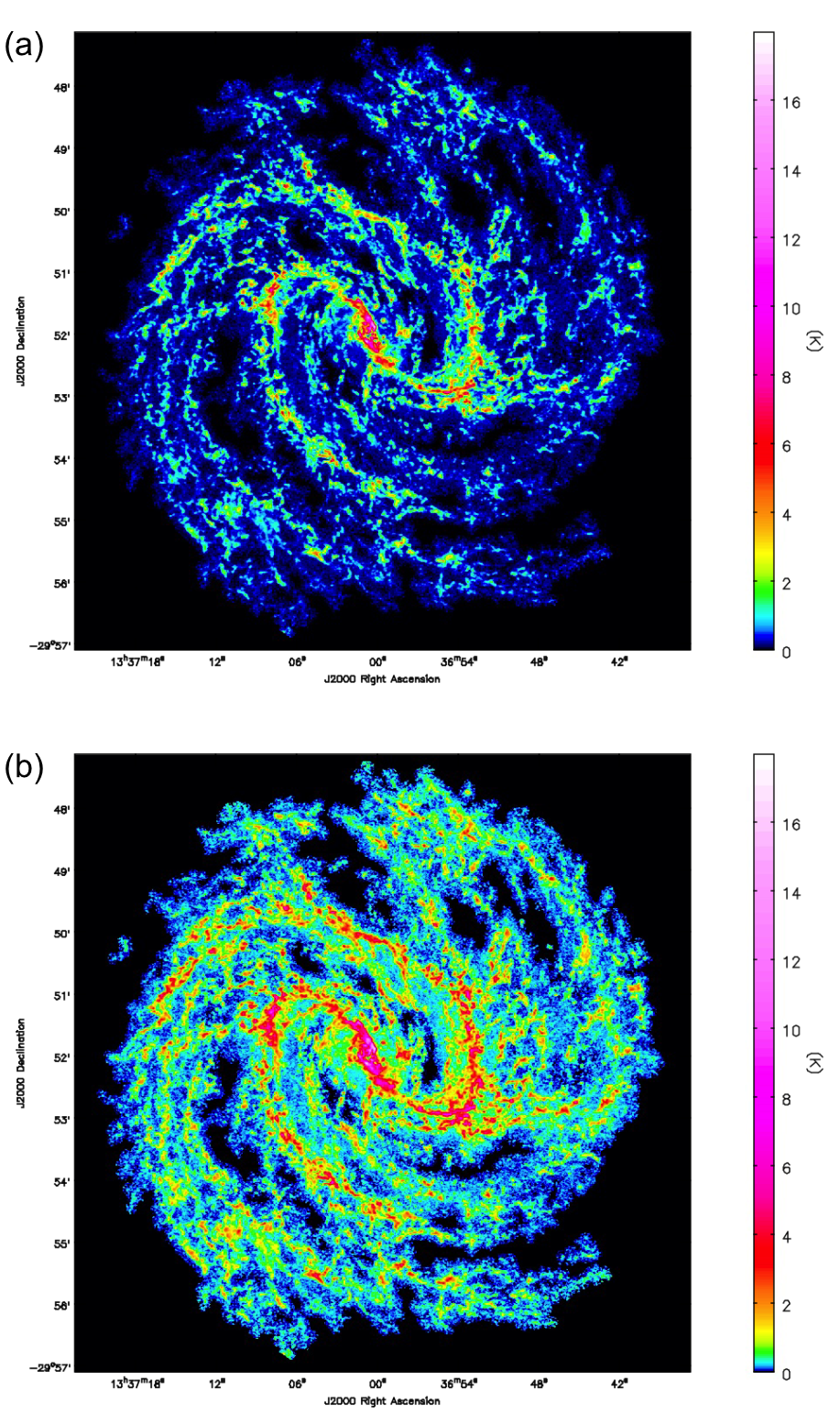}
\caption{
The peak temperature $T_{\rm peak}$ maps. $T_{\rm peak}$ is the maximum
temperature/intensity value among all velocity channels at each spatial pixel.
Two panels, (a) and (b), show the same map, but with different color scales.
Panel (b) shows the low $T_{\rm peak}$ component clearer.
\label{fig:combmom8}}
\end{figure*}

\begin{figure*}[h]
\epsscale{0.85}
\plotone{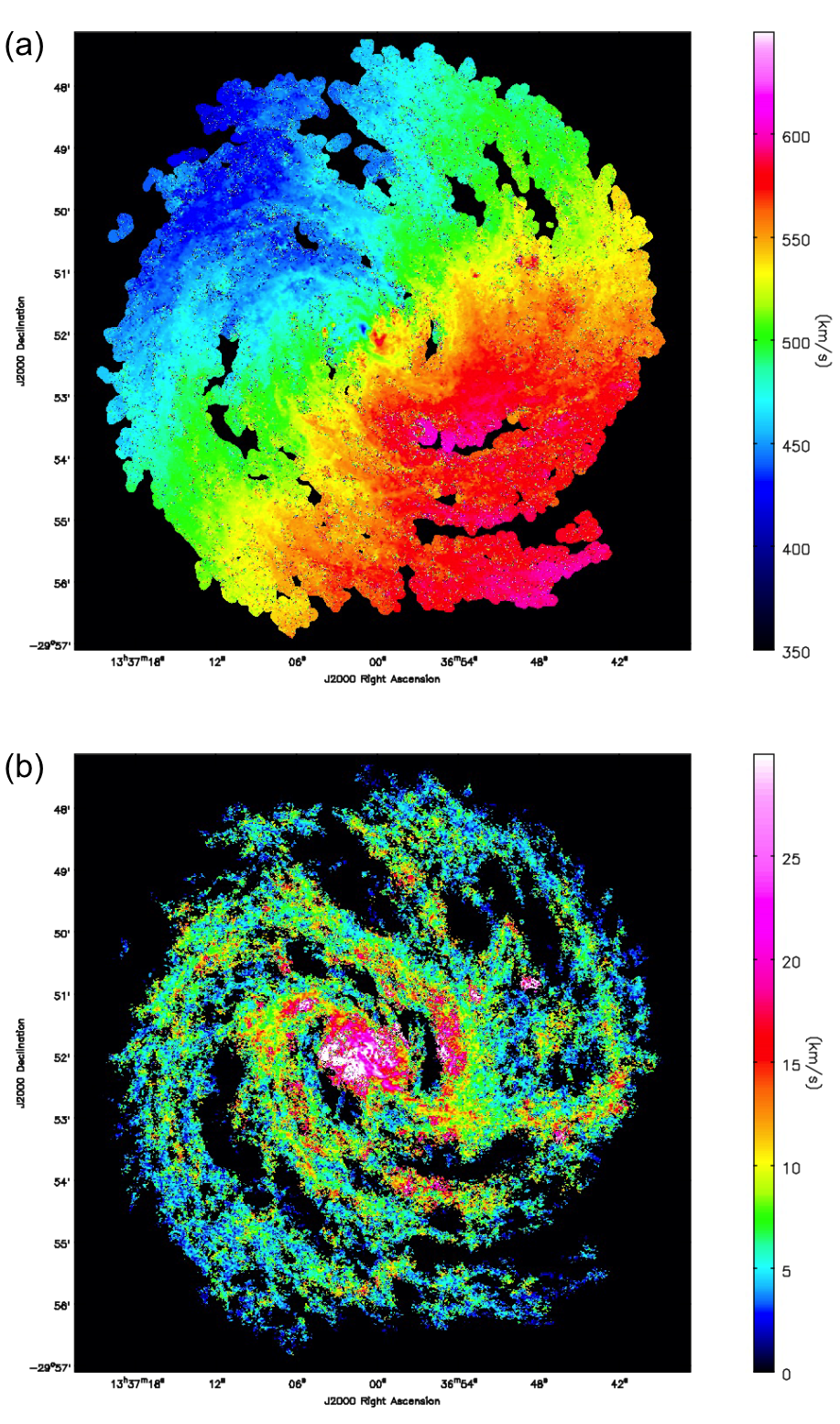}
\caption{
(a) The velocity field $V_{\rm los}$ map (the ``mom1" map).
(b) The velocity dispersion $\sigma_{v}$ map (the ``mom2" map).
Panel (a) is made from the data cube with a $5\kmps$ channel width, while panel (b) is from the cube with a $1\kmps$ channel width to measure $\sigma_{v}$ smaller than $5\kmps$. 
\label{fig:combmom1mom2}
}
\end{figure*}

Figures \ref{fig:combmom8} and \ref{fig:combmom1mom2}
show the peak/maximum temperature $\Tp$ map,
line-of-sight velocity field $V_{\rm los}$ (mom1) map, and
the velocity dispersion $\sigma_v$ (mom2) map, respectively.
The mask is applied in making the $I_{\rm CO}$, $V_{\rm los}$, $\sigma_v$ maps, but not for the $\Tp$ map.
We use the cube with the $5\kmps$ channel width for generating these maps
except for the $\sigma_v$ map, where
we use the $1\kmps$ cube so that dispersions narrower than $5\kmps$ can be measured.
The mask from the $5\kmps$ cube is regridded and used to produce the $\sigma_v$ map.
 
The noise level $\sigma_{I_{\rm CO}}$ varies across the $I_{\rm CO}$ map
(Figure \ref{fig:combmom0})
depending on how many pixels along the velocity axis are included for the velocity integration.
Specifically, $\sigma_{I_{\rm CO}}=({\rm RMS} \cdot {\Delta} V) N^{1/2}$,
where RMS is the rms for channel width ${\Delta}V$ and $N$ is the number of integrated spectral channels. 
As a guideline, the RMS noise of 96~mK in a $5\kmps$ channel
translates to $0.48\,\rm K\kmps$.
Thereby, most emission evident to eyes in Figure \ref{fig:combmom0}
is detected at a high significance.
The maximum integrated intensity is $1023 \,\rm K\kmps$.

\begin{figure}[h]
\epsscale{1.1}
\plotone{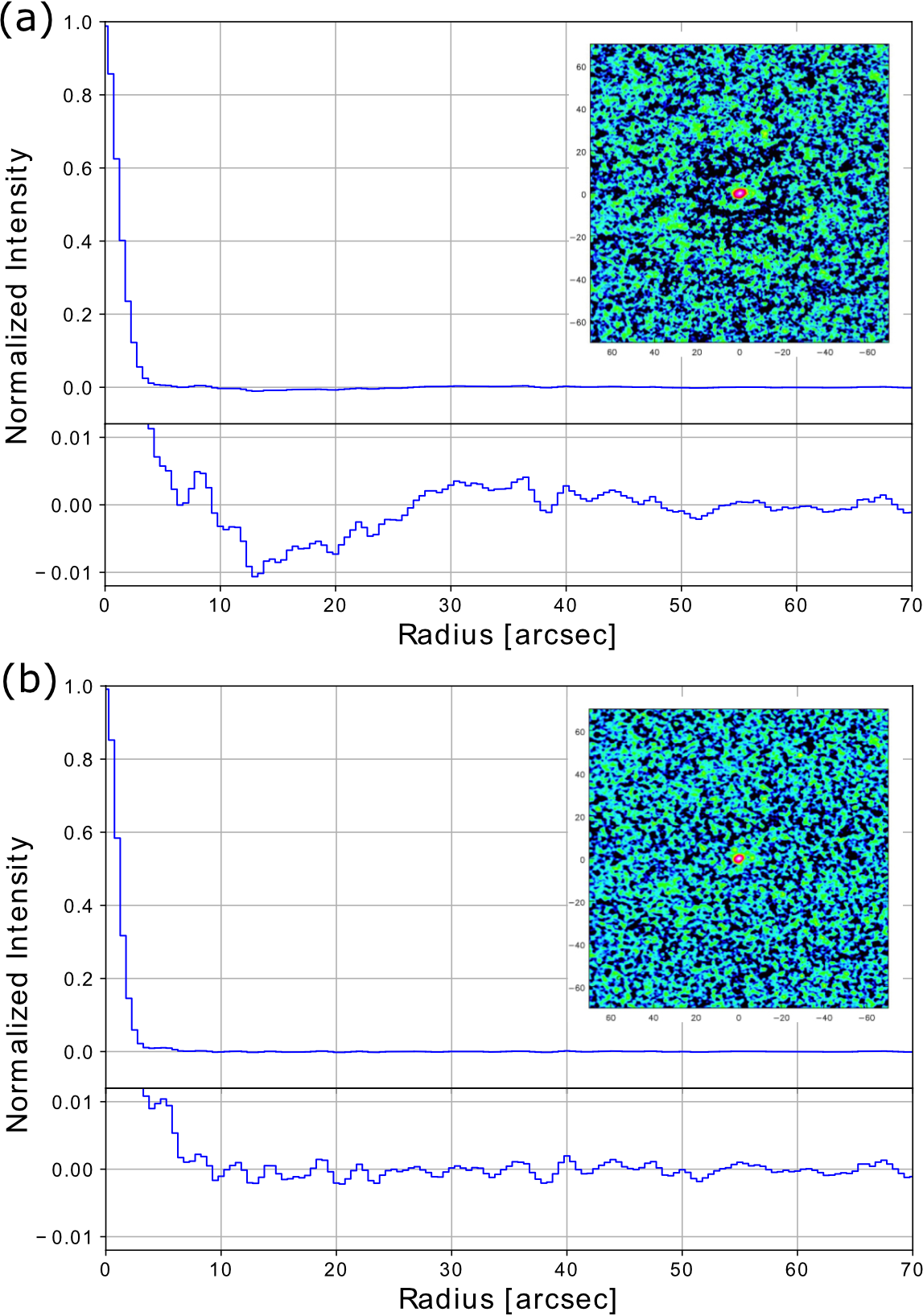}
\caption{
Normalized emission profiles of an isolated compact source,
for an evaluation of sidelobes on large scales:
(a) for the dirty map before CLEAN, and
(b) for the CLEANed map.
The peak intensity is scaled to 1.
Each panel consists of a radial profile (top) and zoom-up of $\pm 1\%$ range (bottom).
A $140\arcsec \times 140\arcsec$ cutout map around the source is the inset.
The source is a high velocity component with
a size of $3.1\arcsec \times 2.1\arcsec$ (P.A.=-66$\arcdeg$)
centered
at ($\alpha_{\rm J2000}$, $\delta_{\rm J2000}$)=($13\fh 37\fm 01.0\fs$, $-29\fd 51\fm 55.4\fs$).
The channels from 350 to 390 $\kmps$ are integrated.
After CLEAN, there is almost no large-scale, systematic sidelobe
to the level of $\sim$0.1\% of the emission peak.
\label{fig:pntprofile}}
\end{figure}

The sidelobes of a PSF often remain after imaging
and are an obstacle for high dynamic imaging,
but this is not the case here.
Figure \ref{fig:pntprofile} shows radial emission
profiles of an isolated, compact source (a) before and
(b) after CLEAN.
This emission is a high velocity wing of a source
at ($\alpha_{\rm J2000}$, $\delta_{\rm J2000}$)=($13\fh 37\fm 01.0\fs$, $-29\fd 51\fm 55.4\fs$)
near the galactic center.
The channels between 350 to 390 $\kmps$ show only this emission component;
these channels are integrated, and the peak intensity is scaled to 1.
The zoom-in (bottom plot) shows that there are
no large-scale, systematic sidelobes left in the CLEANed image (panel b)
down to $\sim 1/1000$.

\section{Molecular Gas Mass and Surface Density} \label{sec:gasmass}

The H$_2$ mass and molecular gas surface density are derived from the measured CO intensities. 
The total integrated CO(1-0) flux over the disk
is $S_{\nu} dv =1.67\times 10^4 \Jykmps$ in the TP cube (\S \ref{sec:redtp})
[$1.68\times 10^4 \Jykmps$ in the 12m+7m+TP cube (\S \ref{sec:imaging})].
This flux is translated to the total H$_2$ gas mass of $\Mmol=2.6\times 10^9 ~\Msun$
and total molecular gas mass of $\Mgas=3.6\times 10^{9}~\Msun$
using the equations \begin{eqnarray}
    \Mmol &=& 7.8\times 10^5 ~\Msun \left(\frac{S_{\nu} dv}{1\,\rm Jy\cdot \kmps}\right)\left(\frac{D}{10\rm \, Mpc}\right)^{2} \nonumber \\
    & & \times  \left(\frac{X_{\rm CO}}{2.0\times 10^{20}\,\rm cm^{-2}\cdot [K\cdot \kmps]^{-1}}\right), \label{eq:H2mass}
\end{eqnarray}
and
\begin{equation}
    \Mgas=1.36\Mmol. \label{eq:gasmass}
\end{equation}
$\Mgas$ includes the masses of helium and other heavier elements
that coexist with H$_2$.
We will analyze the CO-to-H$_2$ conversion factor with this data in the future.
For this paper, we temporarily
adopt the consensus value of $X_{\rm CO}=2.0\times 10^{20} \,\rm cm^{-2}\cdot [K\kmps]^{-1}$
\citep{Bolatto:2013ys}
for simplicity,
with caveats that this value has at least a factor of two uncertainty \citep{Bolatto:2013ys},
may vary with galactic radius \citep{Lada:2020aa},
and some of the measurements summarized in \citet{Bolatto:2013ys} are
questioned \citep{Koda:2016aa, Scoville:2022aa}.
The optical isophotal diameter of the galaxy is $D_{25}=12.9\arcmin$ (16.9~kpc).
The average molecular gas and total surface densities over the optical disk are
$\Sigmamol^{\rm tot} \sim 12\,~\Msun /\pc^2$ and
$\Sigmagas^{\rm tot} \sim 16\,~\Msun /\pc^2$.

The $\Mmol$ estimated here is consistent with the previous measurements:
practically the same as $2.5\times 10^9 ~\Msun$ by \citet{Crosthwaite:2002yu}
and 26\% lower than $3.4\times 10^9 ~\Msun$ by \citet{Lundgren:2004aa}.
For the latter, we re-calculated the mass using the $X_{\rm CO}$ adopted in our analysis.
The discrepancy is not surprising since the telescopes, instruments, and weather for this previous study were not as stable as the observing environment at ALMA.

The CO(1-0) integrated intensity $I_{\rm CO}$
is translated to the molecular gas surface density as
\begin{eqnarray}
    \Sigmamol &=& 3.2 \, ~\Msun / \pc^2 \cos(i) \left(\frac{I_{\rm CO}}{\rm K \kmps}\right) \nonumber \\
    & & \times  \left(\frac{X_{\rm CO}}{2.0\times 10^{20}\,\rm cm^{-2}\cdot [K\cdot \kmps]^{-1}}\right), \label{eq:H2surfdens}
\end{eqnarray}
and the total gas surface density, by including helium and other heavier elements, as
\begin{equation}
    \Sigmagas = 1.36\Sigmamol. \label{eq:gassurfdens}
\end{equation}
The $i$ is an inclination of the disk.

Under the assumption of a constant $X_{\rm CO}$,
Figure \ref{fig:combmom0} can be seen as the $\Sigmamol$ or $\Sigmagas$
distributions by multiplying 2.9 or 3.9,
the coefficients of eqs. (\ref{eq:H2surfdens}) and (\ref{eq:gassurfdens}) when $i=26\arcdeg$ (derived in Section \ref{sec:kinparam}).
The RMS noise of $0.48\,\rm K\kmps$ in each channel is $\Sigmagas\sim 1.9~\Msun / \pc^2$.
The maximum integrated intensity of $1023\,\rm K\kmps$,
at the inner edge of the northern offset ridge,
is $\Sigmagas\sim 4000~\Msun / \pc^2$.

\begin{figure}[h]
\epsscale{1.2}
\plotone{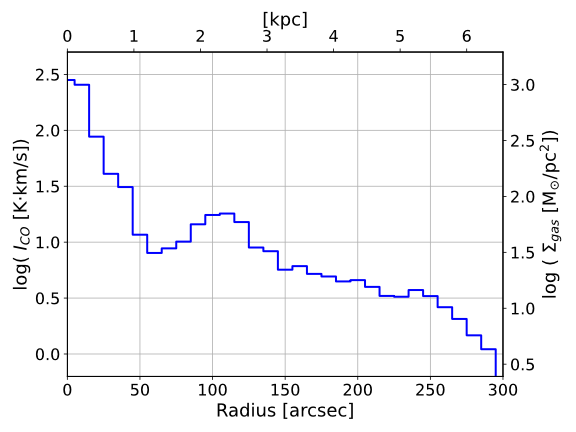}
\caption{
Radial profile of integrated intensity and gas surface density.
The azimuthal average is calculated in each $10\arcsec$ bin from the galactic center.
The top and right axes are calculated assuming the distance to the galaxy,
and with the CO-to-H$_2$ conversion factor including He and other heavy elements.
The adopted position angle and inclination are listed in Table \ref{tab:m83}.
\label{fig:profile}}
\end{figure}

Figure \ref{fig:profile} shows the radial profile of $I_{\rm CO}$ and $\Sigmagas$.
The azimuthal average is calculated
in each $10\arcsec$ ($218\pc$) bin, including the corrections for
the inclination and position angle of the disk.
It shows an overall declining trend from the center to the disk outskirts,
with clear concentrations at the galactic center and 
around the bar end ($r\sim100\arcsec$).

\section{Kinematic Parameters} \label{sec:kinparam}
Figure \ref{fig:combmom1mom2}a shows that
the molecular gas motion is predominantly governed by galactic rotation,
with additional local perturbations due to galactic structures, such as the bar and spiral arms.
We apply the 3D-Barolo (BBarolo) tool \citep{Di-Teodoro:2015aa}
to the data cube and derive the kinematic and geometric parameters of the rotating disk
[i.e., \ systemic velocity ($V_{\rm sys}$), rotation velocity ($V_{\rm rot}$),
position angle (PA), and inclination angle ($i$)]
for the rest of analysis.
The derived parameters are summarized in Table \ref{tab:m83}.
A more thorough analysis of disk kinematics will be presented elsewhere.

For stability in the fitting, we derive the parameters in three steps.
Since we need only the global disk parameters,
we adopt a $10\arcsec$ radial bin to speed up the fitting.

First, we fix the center position to the symmetry center of $K$-band isophotes
\citep[$\alpha_{\rm J2000}=13\fh37\fm0.57\fs$ and $\delta_{\rm J2000}=-29\fd51\fm56.9\fs$; ][]{Thatte:2000aa, Diaz:2006aa}
and obtain $V_{\rm sys} = 511 \pm 3\kmps$ for the radial range of $r = 20$--$280\arcsec$.
We exclude the central region from the fit since it shows exceedingly high velocity components,
which are unlikely related to the galaxy's global dynamics.

Second, we fix ($\alpha_{\rm J2000}$, $\delta_{\rm J2000}$) and $V_{\rm sys}$,
and derive PA $= 225 \pm 1 \arcdeg$ and $i = 26 \pm 2 \arcdeg$
for the radial range of $r = 100$--$250\arcsec$ where the rotation curve is almost flat.
The $r \lesssim 100\arcsec$ area is excluded, because 
the bar ends are around $r \sim 100\arcsec$ and non-circular motions are significant.

Third, the final rotational velocity is derived with all these parameters fixed.
The average value in $r\gtrsim 100\arcsec$
is $V_{\rm rot} = 174 \pm 10\kmps$, or $\Delta V=348\kmps$ in the full width.
These parameters are consistent with the ones derived
in the previous CO study \citep{Crosthwaite:2002yu}.

We adopt the pattern speed of the bar $\Omega_{\rm b}=57.4\kmps \kpc^{-1}$
from \citet{Hirota:2014wd}, which is derived from an analysis of
the geometry and kinematics of the gas in the bar offset ridges.
The corotation radius of the bar is
$R_{\rm CR}=V_{\rm rot}/\Omega_{\rm b}=3.0\kpc$ ($2.3\arcmin$).
It is not clear if the spiral arms have a constant pattern speed (see Section \ref{sec:dynamicevolution}).

\section{Molecular Gas Distribution} \label{sec:distribution}

The integrated intensity and peak temperature maps (Figures \ref{fig:combmom0}-\ref{fig:combmom8})
show numerous local peaks across the disk,
which correspond to molecular clouds
or their associations \citep[giant molecular cloud associations - GMAs; ][]{Vogel:1988tp}.
They also show some of the common large-scale molecular structures
among barred spiral galaxies, including
the concentration of molecular gas at the galactic center,
narrow offset ridges along the bar,
concentrations around the bar ends, and
molecular spiral arms extending from the bar ends to the disk outer part.
Figures \ref{fig:combmom0}-\ref{fig:combmom8} also show
other molecular structures, such as
diffuse extended emission around the bar and spiral arms,
filamentary structures in the interarm regions,
and bifurcation or multiple branches of the molecular spiral arms
toward the disk outskirts.

\begin{figure}[h]
\epsscale{1.1}
\plotone{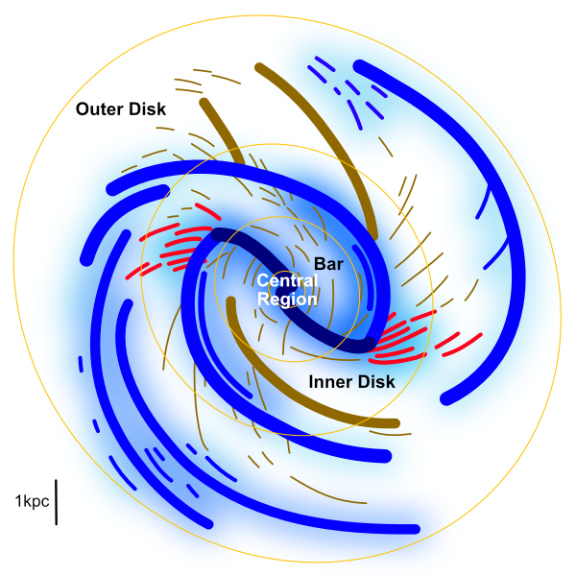}
\caption{
Schematic illustration of molecular gas structures, drawn based on Figures \ref{fig:combmom0} and \ref{fig:combmom0alternative}.
The orange ellipses enclose the four regions, the central region ($<20\arcsec$), bar (20-80$\arcsec$), inner disk (80-160$\arcsec$), and outer disk (160-300$\arcsec$).
The blue lines trace the bar and spiral arm structures with the line thicknesses roughly indicating the significance of emission.
The blue shade shows the extended gas distributions with darker and lighter blues for prominent and less prominent emission, respectively.
The red and brown lines are structures in the interarm regions,
and the red-marked are the ones around the transitions from the bar ends to spiral arms.
This figure does not aim to identify every structure in M83.
\label{fig:schematic}}
\end{figure}

\begin{figure*}[h]
\epsscale{1.1}
\plotone{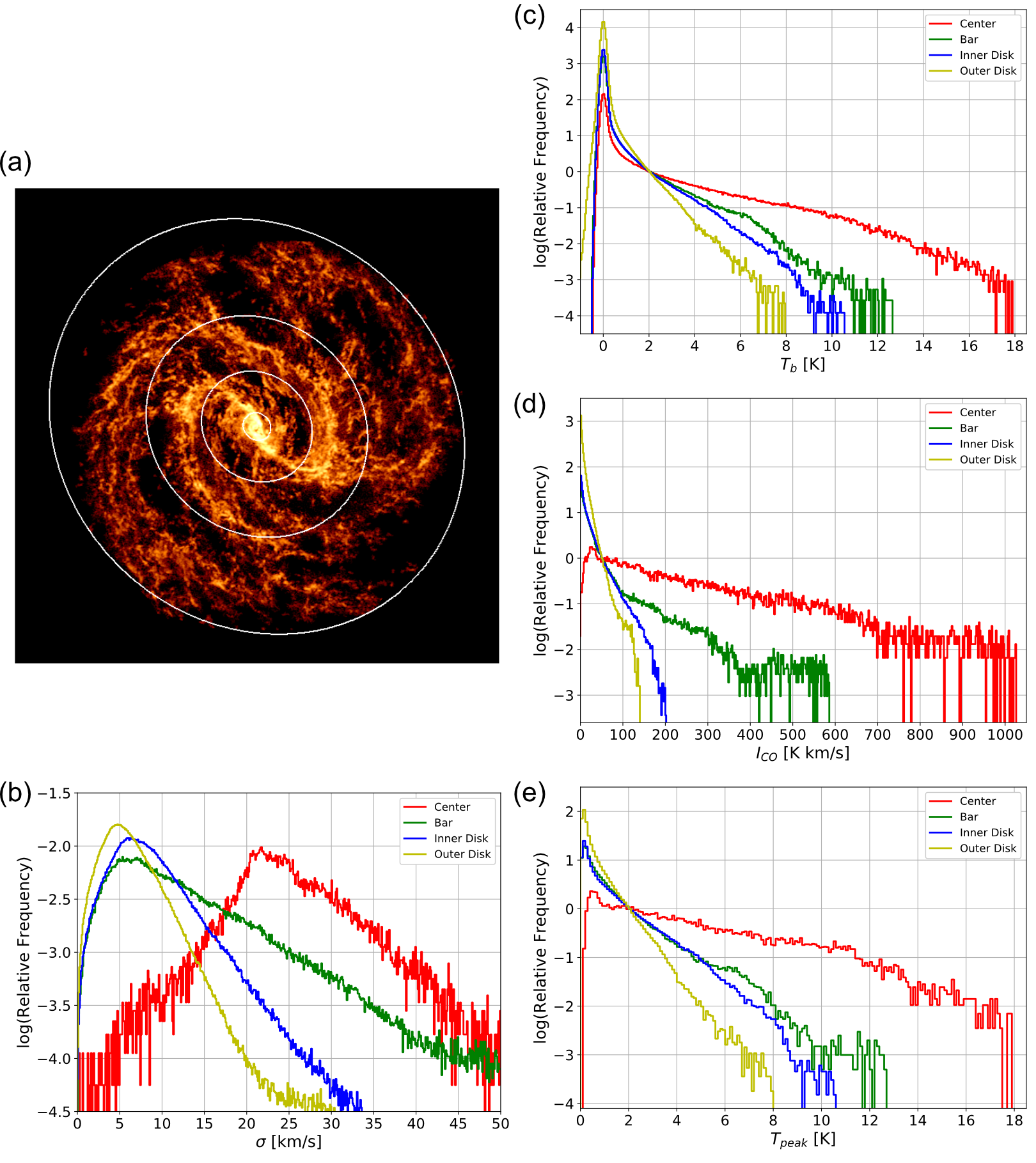}
\caption{
Probability distribution functions (PDFs).
(a) Definition of circular regions around the galactic center:
the center ($r<20\arcsec$),
bar (20-80$\arcsec$),
inner disk (80-160$\arcsec$),
and outer disk regions (160-300$\arcsec$).
The 20, 80, 160, and 300$\arcsec$ correspond to 0.44, 1.75, 3.49, and 6.54~kpc, respectively.
The remaining panels show the PDFs for
(b) velocity dispersion $\sigma$,
(c) brightness temperature $\Tb$ in the data cube,
(d) integrated intensity $I_{\rm CO}$, and
(e) peak brightness temperature $\Tp$.
The PDFs are made with the data of (b) Figure \ref{fig:combmom1mom2}b, (c) the $5\kmps$ cube, (d) Figure \ref{fig:combmom0}, and (e) Figure \ref{fig:combmom8}.
The $y$ axis of each plot is a relative frequency, proportional to the number of pixels.
The $y$ scale is normalized,
so that the sum of the frequencies is one for (b),
and that the frequency is one (c) at $\Tb=2$~K, (d) $I_{\rm CO}=50\,\rm K\cdot \kmps$, and (e) $\Tp=2$~K.
\label{fig:pdfs}}
\end{figure*}

In the following subsections, 
we will discuss the distribution and properties of molecular gas,
by separating the disk into four regions according to their radii in the disk plane ($i=26\arcdeg$):
the central region ($r<20\arcsec$),
the bar region (20-80$\arcsec$),
the inner disk (80-160$\arcsec$),
and the outer disk (160-300$\arcsec$).
These definitions are adapted for the purpose of discussions in this paper,
but not for rigorous classification.
The regions are indicated in Figures \ref{fig:schematic} and \ref{fig:pdfs}a.
Figure \ref{fig:schematic} is a schematic illustration of some molecular structures
discussed in the following subsections.
Figure \ref{fig:pdfs} shows 
the probability distribution functions (PDFs) of
(b) velocity dispersion $\sigma_v$,
(c) brightness temperature in the cube $\Tb$,
(d) integrated intensity $I_{\rm CO}$  (=$\int \Tb  dv$),
and (e) peak brightness temperature $\Tp$ (=$\max \Tb$),
in these regions.

\subsection{Molecular Clouds in the MW Disk as Reference} \label{sec:gmcreference}

\begin{figure}[h]
\epsscale{1.0}
\plotone{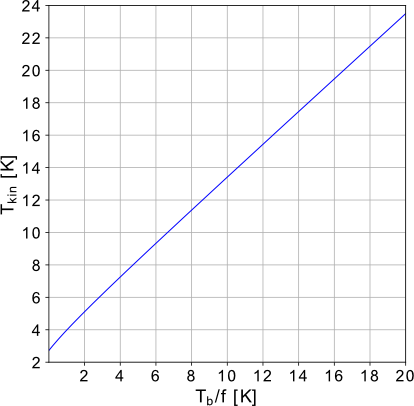}
\caption{
Relation between the Rayleigh-Jeans brightness temperature $\Tb$ and
kinetic temperature $\Tk$ at the CO(1-0) frequency.
The horizontal axis is expressed with $\Tb$ divided by the beam filling factor $f$.
\label{fig:tbtkin}}
\end{figure}

The parameters of typical molecular clouds from CO surveys in the MW are a good reference for our discussion.
Table \ref{tab:gmcmw} lists the mass-weighted average parameters of clouds
within the solar circle from the Massachusetts-Stony Brook Galactic Plane CO(1-0) survey \citep{Scoville:1987vo}.
A typical molecular cloud has $D\sim40\pc$, $\sigma\sim3.8\kmps$ ($8.9\kmps$ in FWHM),
$\Mgas\sim 4\times 10^5~\Msun$, and $\Tk\sim 10\,\rm K$
($\Tp\sim 7\,\rm K$ from Figure \ref{fig:tbtkin} which is presented later).
Newer measurements using the $^{13}$CO emission provide similar cloud parameters \citep{Roman-Duval:2010fk, Koda:2006ab}.
We note that these mass-weighted averages are skewed
toward large and massive clouds.
The clouds of $D\gtrsim 40\pc$ and $\Mgas\gtrsim 4\times 10^5~\Msun$
are rare as a population, but contain half the molecular gas mass in the MW (inside the solar circle).
Most of the clouds in the Solar neighborhood have $\lesssim 10^5~\Msun$
\citep[e.g., the ``Taurus" cloud contains $3\times 10^4~\Msun$; ][]{Dame:1987aa}.
The gas surface densities are roughly constant $\sim 200\,\rm ~\Msun / \pc^2$
among the clouds in the inner Galactic disk, independent of their masses and sizes
\citep{Solomon:1987pr, Heyer:2015qy},
though this measure may decrease radially;
e.g., $\sim 1800\,\rm ~\Msun / \pc^2$ in the Galactic center \citep{Oka:2001qt}
and $\sim 30\,\rm ~\Msun / \pc^2$ in the outer disk \citep{Heyer:2001aa}.

\begin{deluxetable}{lll}
\tablecaption{Parameters of a typical molecular cloud in the MW from \citet{Scoville:1987vo} \label{tab:gmcmw}}
\tablehead{
\multicolumn{2}{c}{Parameter} & \colhead{Value}}
\startdata
  Diameter          & $D$           & $40\pc$  \\
  Mass(H$_2$+He)    & $\Mgas$       & $4\times 10^5~\Msun$  \\
  Density           & $n_{\rm H_2}$ & $180\,\rm cm^{-3}$  \\
  Kinetic Temperature & $\Tk$       & $10\,\rm K$ \\
  Thermal Pressure  & $P_{\rm TH}/k$    & $2000\,\rm cm^{-3} K$ \\
  Velocity Dispersion & $\sigma$    & $3.8\,\rm \kmps$
\enddata
\end{deluxetable}

Our observations of M83 can resolve the typical clouds with
$\Mgas = 4\times 10^5~\Msun$ and $D=40~\pc$
at a spatial resolution of $40~\pc$ and $1\sigma$-sensitivities of 0.096~K in brightness,
$\sim 1.9~\Msun / \pc^2$ in gas surface density,
and $\sim 4.1\times 10^3~\Msun$ in gas mass, in a $5~\kmps$ channel.
We can further detect, but not spatially resolve,
less massive clouds of $\sim 10^4~\Msun$ (at $3\sigma$)
if they are isolated from surrounding clouds
in space and/or velocity.
The typical surface density of molecular clouds in the MW of $\Sigmagas\sim 200 ~\Msun / \pc^2$
corresponds to $I_{\rm CO}\sim 45 \,\rm K \kmps$ (see Figures \ref{fig:combmom0} and \ref{fig:combmom0alternative}).

Most CO(1-0) emission is from optically-thick regions (see Section \ref{sec:COexcitation}).
Under an assumption of thermalized, optically-thick CO(1-0) emission,
$\Tb$, $\Tp$, and $I_{\rm CO}$ are related to the kinetic temperature $\Tk$ as
\begin{equation}
\Tb=f (F(\Tk) -F(T_{\rm CMB})),\label{eq:TkinTb}
\end{equation}
where
\begin{equation}
F(T) = T \left[ \frac{h\nu / k T}{\exp \left( h\nu / k T \right) - 1} \right].
\end{equation}
$T_{\rm CMB} = 2.725\,\rm K$ is the temperature of
the Cosmic Microwave Background (CMB) radiation,
and $h\nu / k = 5.53\,\rm K$ at the CO(1-0) frequency.
$f$ is a beam filling factor, and $f=1$ when the CO emission
occupies the full area of a 40~pc beam aperture.
A single, unresolved, round cloud with a diameter of 20~pc within a beam
would have $f=0.25$ and show $\Tb\sim 1.7\,\rm K$
if its kinetic temperature is $\Tk\sim10\,\rm K$.
Figure \ref{fig:tbtkin} shows the relation between $\Tb$ and $\Tk$ (eq. \ref{eq:TkinTb}).

\subsection{The Central Region: $r<20\arcsec$ ($\sim 0.44\kpc$)} \label{sec:center}

The radial profile of $I_{\rm CO}$ (and $\Sigmagas$)
shows a significant concentration of molecular gas in the central region (Figure \ref{fig:profile}).
This is often observed in barred spiral galaxies \citep[e.g., ][]{Sakamoto:1999qf, Sheth:2005qy, Querejeta:2021aa}
as the bar's elongated potential steers gas motions toward the central region
\citep{Matsuda:1977aa, Simkin:1980aa, Combes:1990aa}.
The central 1-kpc diameter region, slightly larger than our definition of the central region,
contains a total H$_2$ mass of $3.9\times10^8~\Msun$, $\sim$15\% of the total in this galaxy,
with an average molecular gas surface density of
$\Sigmamol^{\rm 1kpc} \sim 497\,~\Msun /\pc^2$.
[Note that for simplicity, we assume a constant $X_{\rm CO}$ across the whole disk,
while it is suggested that 
the $X_{\rm CO}$ in the central region could be %3-10 times
lower than the adopted value
\citep{Sodroski:1995aa, Arimoto:1996aa, Oka:1998vn, Strong:2004aa}.]
The $\Sigmamol^{\rm 1kpc}$ is greater than the average surface density of typical clouds in the inner MW disk.
The gas concentration factor, that is, the excess in surface density compared to the disk average,
is $\Sigmamol^{\rm 1kpc}/\Sigmamol^{\rm tot} \sim 43$.
This ratio is in the range of barred spiral galaxies studied by \citet{Sakamoto:1999qf} and \citet{Sheth:2005qy}
who also used a constant $X_{\rm CO}$ value comparable to this study.

The red lines in Figure \ref{fig:pdfs} show the PDFs of the central region.
The excesses in $I_{\rm CO}$, $\Tb$, and $\Tp$ toward the high values are evident,
and are consistent with the results by \citet{Egusa:2018aa} who studied a smaller region of this galaxy.
The high $\Sigmamol^{\rm 1kpc}$ values likely suggest that the molecular gas occupies the entire beam.
Hence, we assume that the beam filling factor is close to $f=1$,
and that the $\Tb$ in this region is a direct measure of
the $\Tk$ of the bulk molecular gas (see Figure \ref{fig:tbtkin}).
Figures \ref{fig:pdfs}c,e show that the molecular gas there
is often warmer than the temperatures in the typical Galactic clouds ($\Tk>10\rm \, K$, i.e., $\Tb>7$~K).
When averaged in a 40pc aperture, the warmest gas 
shows $\Tk\sim 19$-$21$~K ($\Tb=\Tp\sim 16$-$18$ K).

The velocity dispersion is also enhanced in the central region (Figure \ref{fig:combmom1mom2}b).
We note that the dispersion here is primarily the component perpendicular to the disk as the galaxy is nearly  face-on.
Figure \ref{fig:pdfs}b shows that a typical dispersion, i.e., the peak of the PDF,
is $\sigma_v \sim 23\kmps$.
This is much higher than those of the other regions (in the following subsections).
The excess of $I_{\rm CO}$ and $\sigma_v$ in the centers of other galaxies,
in comparison to their disks, is also found in other barred spiral galaxies \citep{Sun:2020tt}.

M83 has a double nucleus: a $K$-band image shows a visible nucleus and symmetry center
\citep[i.e., the bulge center; ][]{Thatte:2000aa, Diaz:2006aa}.
The symmetry center coincides with the dynamical center \citep{Sakamoto:2004aa}.
The highest $\sigma_v \sim 50\kmps$ is
at the location of the visible nucleus at (R.A., DEC)$_{\rm J2000}$=(13:37:0.95, -29:51:55.5),
which is possibly due to another gas disk rotating around this nucleus \citep{Sakamoto:2004aa}.

\subsection{The Bar: $r=$20-80$\arcsec$ ($\sim 0.44$-$1.8\kpc$)} \label{sec:bar}

The most prominent features in this region are the sharp ridges of emission along the leading sides
of the stellar bar, known as ``offset ridges" \citep[dark blue lines in Figure \ref{fig:schematic}; ][ as early studies]{Ishizuki:1990aa, Kenney:1992aa}.
These molecular ridges are coherent over a radius of  $\sim 2$~kpc and are slightly curved along the bar.
Their widths are as narrow as $\sim$100-200 pc with occasional wiggles with an amplitude of $\sim$ 100-200 pc.
Their surface density is greater than the typical cloud value,
$\Sigmagas\gtrsim 200~\Msun / \pc^2$ ($I_{\rm CO}\gtrsim 45~ \,\rm K \kmps$).
This is also seen in the PDF as a high $I_{\rm CO}$ tail (green in Figure \ref{fig:pdfs}d)
compared to that of the inner disk (blue).
The offset ridges of barred spiral galaxies exhibit little star formation
\citep[e.g., ][]{Downes:1996tk, Momose:2010cr, Maeda:2020aa},
but those in M83 show some associated H$\alpha$ emission indicative of recent star formation activity 
\citep[Figure \ref{fig:hstco}; see also ][]{Hirota:2014wd}.

Despite the excess in $I_{\rm CO}$ in the offset ridges,
the $\Tb$ and $\Tp$ images show only subtle, albeit definite, differences
from those of the molecular spiral arms in the inner disk (Section \ref{sec:innerdisk}).
The excess in $I_{\rm CO}$ originates almost entirely from the excess in
$\sigma_{v}$ along the ridges 
\citep[Figure \ref{fig:pdfs}b; note approximately $I_{\rm CO}\sim \Tb\sigma_{v}$, see also ][]{Egusa:2018aa, Sun:2018aa}.
In other words, the gas temperature is about the same between the bar and inner disk regions
even though the velocity dispersion is enhanced in the bar offset ridges.
The shape of the PDF of $\sigma_{v}$ (Figure \ref{fig:pdfs}b)
appears to show a radial transition from the central region (red) to the inner disk (blue).
The high $\sigma_v$ in the central region is also seen in the bar region (green),
and then, is reduced in the inner disk region in Figures \ref{fig:pdfs}b and \ref{fig:combmom1mom2}b.
The bar region may carry kinematic traits of the central region.
The maximum $\Tb$ in the bar region is
$\sim$10-12~K (Figure \ref{fig:pdfs}c,e) and occurs on the side closer
to the central region (Figure \ref{fig:combmom8}).

This region also shows many filamentary structures
outside the offset ridges (Figures \ref{fig:combmom0} and \ref{fig:combmom8}; thin brown lines in the bar region in Figure \ref{fig:schematic}).
They are often 1~kpc and sometimes longer in size, and some are connected to the offset ridges almost perpendicularly
on their upstream sides assuming that the disk is rotating clockwise.

Extended, low-brightness CO emission around the offset ridges
are also seen in Figures \ref{fig:combmom0} and \ref{fig:combmom0alternative}
(typically $\sim 10 \,\rm K \kmps$; $\sim 10^4~\Msun$ in each 40pc beam; blue shade in Figure \ref{fig:schematic}).
Figure \ref{fig:combmom0alternative} shows where $\Sigmagas$ exceeds
the average cloud surface density ($\sim 45 \,\rm K \kmps$) and where it is below,
and hence, shows the low-brightness emission surrounds the offset ridges.
While the presence of such extended emission was inferred
in an ALMA study of the barred galaxy NGC 1300 
to explain a missing flux in the interferometer data \citep{Maeda:2020aa}, 
this is the first study that such emission is clearly imaged around a bar.
The peak temperature map (Figure \ref{fig:combmom8}) shows that
bright peaks and ridges/filaments are spatially localized and embedded in
the low-brightness emission of much larger extents.
These CO structures show only little star formation
-- the extended CO emission exhibits almost no H$\alpha$ emission,
and the filaments show little H$\alpha$ emission (Figure \ref{fig:hstco}).
The $\sigma_{\rm v}$ is elevated around these emission 
and often reaches $\sim 20$-$40\kmps$ (Figure \ref{fig:combmom1mom2}b).

Small areas around the extended, low-brightness emission show
virtually no CO emission, indicating that there is no molecular gas down to 
$\sim 10^4~\Msun$ within the 40~pc aperture.
These areas can be more clearly identified in Figure \ref{fig:combmom1mom2}a
as they appear as voids (black) in the mask (Section \ref{sec:maps}).

\subsection{The Inner Disk: $r=$80-160$\arcsec$ ($\sim 1.8$-$3.5\kpc$)} \label{sec:innerdisk}

The definition of the inner disk includes the bar ends and
inner sections of the two spiral arms (dark blue and blue lines in Figure \ref{fig:schematic}).
The radial profile  (Figure \ref{fig:profile}) shows
a notable bump around $r\sim 100\arcsec$ (2.2~kpc),
which corresponds to the transition regions
from the bar ends to the beginning of the two main spiral arms (Figure \ref{fig:combmom0}).
The bump includes emission from the spiral arms,
as they run almost at a constant galactic radius,
as well as the emission from the gas concentrations at the bar ends.
Such bar-end concentrations are often observed in barred spiral galaxies \citep{Sheth:2002lr}.
Compared to the long offset ridges along the bar,
the gas distribution appears more fragmented into
shorter filaments and small blobs, an ensemble of which form
the concentrations at the bar ends and the spiral arm structures.
The peak brightness temperatures in the inner disk region are
$\sim$8-10~K (Figure \ref{fig:pdfs}c,e) and occur primarily
in these concentrations (Figure \ref{fig:combmom8}).

\begin{figure*}[h]
\epsscale{1.1}
\plotone{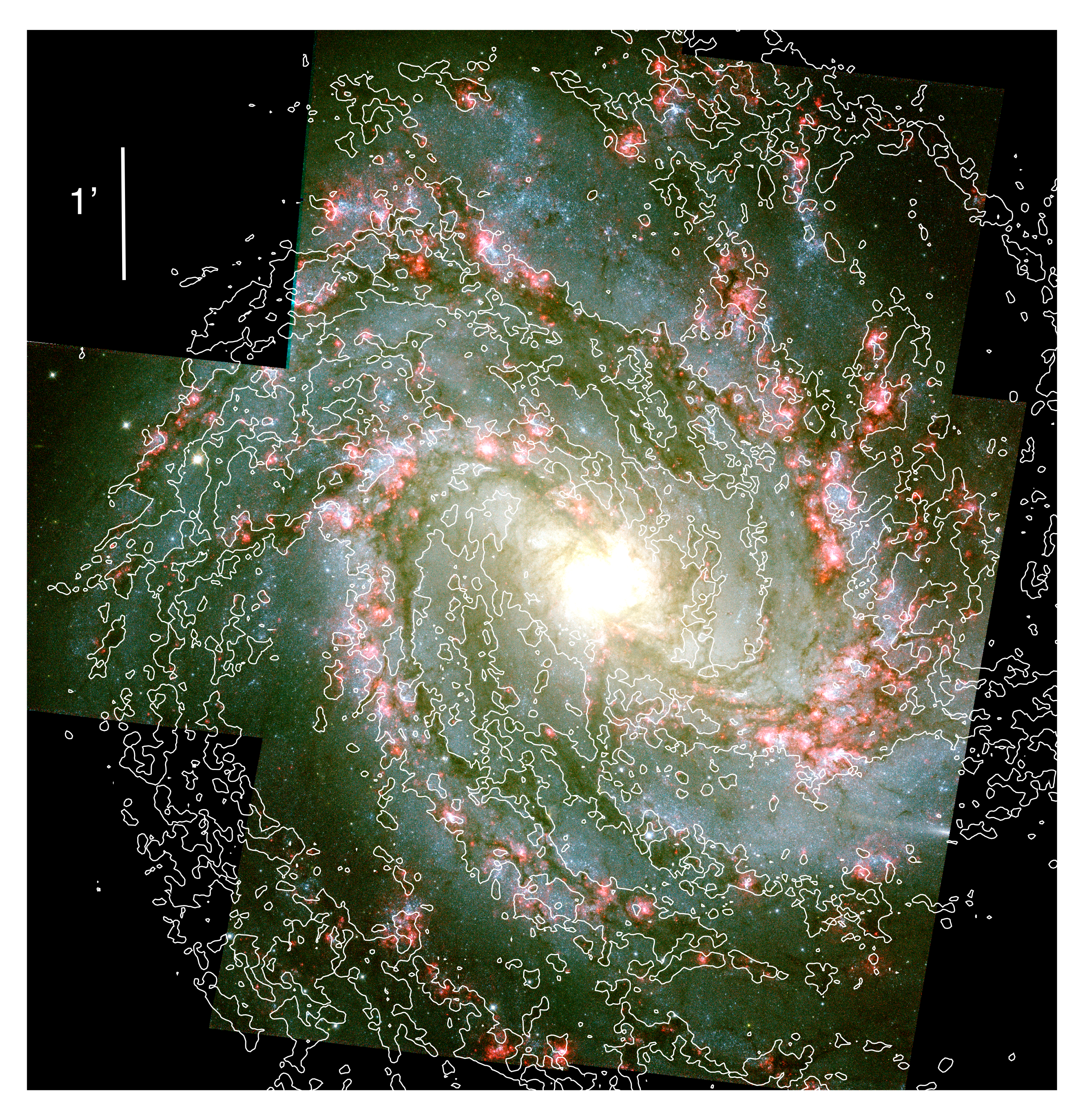}
\caption{
A CO(1-0) integrated intensity contour on a pseudo-color (RGB) image from the Hubble Space Telescope.
The CO contour is at 5 $\rm K \kmps$.
The HST image is made with the F657N (H$_\alpha$) image for R, a geometric mean of the F438W and F814W images (i.e., $\sqrt{\rm F438W \times F814W}$) for G, and the F438W image for B, all of which are taken from \citet{Blair:2014aa}.
\label{fig:hstco}}
\end{figure*}

The gas concentrations at the bar ends transition into two major molecular spiral arms
with a width of $\sim$$30\arcsec$ ($\sim 600\pc$, including only visually-distinct bright parts of the arms).
These spiral arms have a small pitch angle and run
at a near constant galactic radius (approximately the corotation radius) in the inner disk region.
In a closer look, these major arms consist of two or more narrower filaments
which run along the directions of the major arms (see Figure \ref{fig:schematic}).
Each of these filaments has a typical width of $\sim$100-200~pc.
The H$\alpha$ emission is associated mainly with the outer filaments (the convex sides),
and not much with the inner filaments
\citep[Figure \ref{fig:hstco}; ][also reported them in their study of a smaller region]{Hirota:2018aa}.

In addition, many shorter, filamentary structures run in/out of
the spiral arms from/into the interarm regions
(Figures \ref{fig:combmom0} and \ref{fig:combmom0alternative}; red and thin brown lines in Figure \ref{fig:schematic}).
These are the structures seen as dust lanes in optical images,
dubbed ``spurs" or ``feathers" \citep{Elmegreen:1980aa, La-Vigne:2006aa, Chandar:2017aa},
and are also observed in CO in other galaxies,
most intensively in M51 \citep{Corder:2008aa, Koda:2009wd, Schinnerer:2017aa}.
Similar filamentary structures are also suggested in the Milky Way 
\citep{Ragan:2014aa, Abreu-Vicente:2016aa, Zucker:2018aa, Veena:2021aa}.
In M83, many of the spurs are associated with HII regions (Figure \ref{fig:hstco}),
which is most evident along the western spiral arm.
The filamentary structures are particularly abundant at the starting parts of
both spiral arms just outside the bar ends (red lines in Figure \ref{fig:schematic}).
Their symmetric occurrence at both arms possibly suggests
large-scale galactic dynamics for generating the molecular structures.

The bright, distinct features discussed so far are surrounded by
fainter, extended CO emission (Figure \ref{fig:combmom0alternative}; blue shade in Figure \ref{fig:schematic}).
In particular, the emission around the western spiral arm,
on its concave side, is
as broad as $\sim 50\arcsec$ ($\sim 1\kpc$)
with a typical average surface density of $\Sigmagas\gtrsim 20 ~\Msun / \pc^2$ ($I_{\rm CO}\gtrsim 5 \,\rm K \kmps$).
The $\sigma_{v}$ is often as high as $\sim 15$-$25\kmps$.

Several long spiral arm-like features are seen in the interarm regions
(outside the major spiral arms; Figure \ref{fig:combmom0}
and thick brown lines in Figure \ref{fig:schematic}).
The most prominent is the one at the southern part,
from at least around the inner radius of the inner disk region
($r\sim 80 \arcsec$ around the seven to eight o'clock direction)
to beyond the outer radius ($r\sim 160 \arcsec$, five o'clock).
The full extent is difficult to trace, but it 
potentially stretches inward into the bar region,
and outward connected to the outer spiral arm running
along the western edge of our field of view.
The length of this feature is at least $190\arcsec$ ($3.8\kpc$)
(only the part within the radial range of the inner disk region)
and possibly longer.
The width increases from about $15\arcsec$ (300~pc)
at the inner radius to $30\arcsec$ (600~pc) at the outer radius.
The dearth of H$\alpha$ emission around this feature is remarkable,
given it is a prominent molecular structure (Figure \ref{fig:hstco}).
Other similar filamentary features in the interarm regions are also evident
in Figure \ref{fig:combmom0}.
Many are long ($\gtrsim 1\kpc$) and narrow (100-200~pc).

A bifurcation of the western spiral arm starts in the inner disk region
at around 2 o'clock (thick blue and brown lines in Figure \ref{fig:schematic}).
The outer, bifurcated branch (brown) extends into the outer disk region
and is as long as $\sim 3.5\arcmin$ (4.6~kpc).
The main arm (blue) is wider
and more massive than the bifurcated arm (outside).
It is notable that H$\alpha$ emission is abundant around the
bifurcated arm, but not as much around the main arm (Figure \ref{fig:hstco}).
On the concave side of the main arm,
there are some short filaments running almost in parallel to, or approaching toward,
the main arm (thin brown lines in Figure \ref{fig:schematic}).
These filaments may collectively form a long spiral arm-like feature in the inter arm region.
This may be a counterpart of the arm-like feature
in the southern interarm region discussed in the previous paragraph.

In addition to these prominent features, there are numerous isolated peaks
(likely individual molecular clouds) in the interarm regions (Figure \ref{fig:combmom0}; not illustrated in Figure \ref{fig:schematic}).
These are more clearly seen in the $\Tp$ map as blobs or dots (Figure \ref{fig:combmom8}).

The PDFs of $\Tb$ and $\Tp$ in the inner disk are similar to
those in the bar (Figure \ref{fig:pdfs}c,e),
suggesting that the physical temperature of the gas is similar between
these regions.
The $\Ico$ in the inner disk, however, does not show the excess
toward high values compared to the bar region (Figure \ref{fig:pdfs}d).
This is because the velocity dispersion $\sigma_{v}$ is overall lower in the inner disk.
The PDF of $\sigma_{v}$ in this region peaks at $\sigma_{v}\sim 6.0\kmps$,
substantially smaller than the one in the central region.

\subsection{The Outer Disk $r=$160-300$\arcsec$  ($\sim 3.5$-$6.6\kpc$)} \label{sec:outerdisk}

Two molecular spiral arms clearly exist in the outer disk region as continuations
of those in the inner disk (Figure \ref{fig:combmom0}).
However, they are broader and less defined than the inner counterparts.
The gas distribution appears more flocculent in the broad spiral arms.
Figure \ref{fig:schematic} illustrates them with blue lines to show their locations,
but the actual emission distribution is more fragmented in Figure \ref{fig:combmom0}.

The eastern/southern arm in this region has a width of 700 pc to 1 kpc 
(blue shaded area in Figure \ref{fig:schematic}).
It appears, in Figure \ref{fig:combmom0}, as a bundle of narrower filaments or ripples,
each of which is about 100-200 pc in width, running almost parallel to each other.
As a whole, these filaments form a molecular spiral arm, which is not as confined
as the molecular arms in the inner disk.
This arm shows only little H$\alpha$ emission, while 
only a small portion is covered in Figure \ref{fig:hstco}.

The western/northern arm runs along the outer boundary of
the outer disk region (Figure \ref{fig:pdfs}a and Figure \ref{fig:schematic}).
It is not clear if this arm is an extension of the eastern arm in the inner disk
or that of the ``interarm" arm discussed in Section \ref{sec:innerdisk}
-- morphologically, it also looks as if these two inner arms merge into this outer-disk arm.
Several filaments run into or out from the main ridge of this arm
but are not aligned with the arm as the filaments in the eastern/southern arm
(e.g., thinner blue lines connected this arm from its concave side in Figure \ref{fig:schematic}). 
The bifurcated branch of the inner western arm discussed in Section \ref{sec:innerdisk} (thick brown line in Figure \ref{fig:schematic})
also merges into this outer arm around the 0 o'clock direction (P.A.$\approx 0\deg$).
In addition to the bifurcated branch,
the feathers or spurs that emerge from the two inner disk arms
(Section \ref{sec:innerdisk}) spread out into the interarm regions in the outer disk (brown in Figure \ref{fig:schematic}).

Some interarm spaces, between the main spiral arms and/or bifurcated spiral arms,
appear almost empty except several small peaks (molecular clouds).
These regions are the blank areas in Figure \ref{fig:combmom1mom2}a,
as they do not show significant emission and thus are masked out
(so they appear blank in the velocity field map).
In particular, the empty area around 0 o'clock is also a void in infrared (little dust),
but is prominent in UV emission \citep[OB associations; ][]{Jarrett:2013wi}.
In fact, the HST image in Figure \ref{fig:hstco} shows almost no H$\alpha$ emission,
but shows abundant blue stars/star clusters.

The $\Ico$, $\Tb$, and $\Tp$
are lower in the outer disk compared to the inner disk (Figures \ref{fig:combmom0} and \ref{fig:combmom8}).
This is quantitatively evident in the PDFs (Figure \ref{fig:pdfs}).
On the 40~pc scale, the molecular gas in the outer disk is either
cooler in temperature and lower in surface density than that in the inner disk,
or resides predominantly in clouds smaller than the 40~pc resolution
(or possibly a combination of these conditions).
Either way, such changes in molecular gas properties from the inner to outer disk
are similar to those in the Milky Way \citep{Heyer:2015qy}.
The maximum $\Tb$ values in the outer disk region are 
as low as 6-8~K and occur in the wide spiral arms (Figure \ref{fig:combmom8}).

The PDF of $\sigma_{v}$ in the outer disk peaks at $\sigma_{v}\sim 4.8\kmps$,
which is lower than that in the inner disk with a peak at $\sim 6.0\kmps$
(Figure \ref{fig:pdfs}).
The difference is subtle, but significantly detected.
These typical dispersions are ubiquitous over the inner and outer disk,
and are highly supersonic (note that the sound speed in H$_2$ gas
is $\sim 0.24\kmps$ at 10~K).

\subsection{Masses of the Filamentary Structures} \label{sec:submass}
The previous subsections do not discuss the masses of the molecular structures much,
to avoid the uncertainty in $X_{\rm CO}$.
Here we summarize the masses
of the filamentary structures (brown, red, and some blue lines in Figure \ref{fig:schematic})
and find that their characteristic masses are $\sim 10^7\Msun$.
The numbers in this section suffer from the uncertainty in $X_{\rm CO}$
(at least a factor of 2).

In the bar region, the filamentary structures have masses of
$\sim 2\times 10^6$ to $3\times 10^7\Msun$.
About half of these structures (longer ones) contain $\gtrsim 10^7\Msun$
(Figure \ref{fig:schematic}).
The inner disk is similar.
They have $\sim 3\times 10^6$ to $3\times 10^7\Msun$.
The 6 most massive have $\gtrsim 2\times 10^7\Msun$ and are among the red lines.
The three long spiral arm-like features, or bifurcated branches of the spiral arms,
in the interarm regions (thick brown lines) are
$\sim 1\times 10^8\Msun$, $8\times 10^7\Msun$, and $3\times 10^7\Msun$
for the southern, north-western, and north-eastern features, respectively.
In the outer disk, each (blue) segment in the flocculent (fragmented) arms has
a mass of $\sim 3$ - $7\times 10^6\Msun$.
The two most massive segments, the most western ones entering the western spiral arm from its concave side,
have masses of $\sim 2\times 10^7\Msun$ (northern one) and $\sim 8\times 10^6\Msun$ (southern).

\section{Molecular Clouds and Extended CO Emission} \label{sec:gmcextended}

Numerous local peaks exist across the disk of M83 (Figures \ref{fig:combmom0} and \ref{fig:combmom8}), each of which likely corresponds to a molecular cloud.
The large molecular structures in the spiral arms and interarm regions
also appear as chains of local peaks/clouds (Figures \ref{fig:combmom0} and \ref{fig:combmom8}).
Not all clouds are resolved, and there is unresolved, extended emission.

A detailed analysis of the internal parameters of individual molecular clouds will be presented in a separate paper (Hirota et al. in preparation).
In Section \ref{sec:COexcitation}, we recall the basics on CO(1-0) excitation to understand the observed emission.
Most CO emission, including the unresolved emission, is likely from molecular clouds
(more accurately, from cloud-like gas concentrations, which we call molecular clouds,
independent of their internal dynamical states, e.g., whether they are bound or not). Exciting CO emission outside of clouds is difficult. 
We discuss radial and azimuthal variations of cloud properties (Section \ref{sec:gmctemp}) and
a cloud-origin of the extended CO emission (Section \ref{sec:extendedCO}).

\subsection{Collisional Excitation and Chemistry of CO} \label{sec:COexcitation}

Our mass sensitivity and spatial resolution of $10^4~\Msun$ ($3\sigma$) and $40\pc$
can detect, but cannot spatially resolve, molecular clouds smaller than a typical one in the MW disk
($\Mgas \sim 4\times 10^5~\Msun$, $D \sim 40\pc$).
In discussing the full CO emission including the unresolved component,
we should recall that the collisional excitation and chemistry of CO
requires volume and column densities similar to those of molecular clouds.

In fact, the average volume density of a typical cloud
($\sim 180 \,\rm cm^{-3}$ in Table \ref{tab:gmcmw}) is nominally
not high enough for collisional excitation of CO with H$_2$.
The critical density is an order of magnitude higher
\citep[$\sim 2,000 \,\rm cm^{-3}$, ][]{Scoville:1987vo}.
Nevertheless, a high optical depth within molecular cloud enables the excitation
as it prevents photons from escaping from the region.
This ``photon trapping" effectively reduces the spontaneous emission rate,
and as a result, decreases the effective critical density
to around the average density within clouds \citep{Scoville:1974yu}.
For this reason, the bulk gas within molecular clouds can emit CO photons even at low volume density.

Paradoxically, for molecular gas outside of molecular clouds -- if it exists -- to emit significant CO emission, it has to have at least the same, or even higher, volume density or column density than that within clouds.
Such conditions are unlikely outside of clouds
except in galactic centers \citep{Oka:1998vn, Sawada:2001lr},
and are unlikely to be ubiquitous across the disk.
Therefore, the great majority of the CO emission from galactic disks should be from clouds.

In addition, there is a requirement from chemistry.
CO molecules are subject to photo-dissociation by the ambient stellar radiation field
(even by the weak field around the Sun, which is not in a major spiral arm).
The presence of CO molecules, and their emission, requires protection
by self-shielding, in addition to dust-shielding, with a sufficient column density, 
as well as replenishment by efficient CO formation in a high volume density \citep{Solomon:1972aa, van-Dishoeck:1988br}.
At a cloud density in a radiation field similar to that in the solar neighborhood, the required column density corresponds to a visual extinction of
$A_{\rm V}\sim $1~mag \citep{van-Dishoeck:1988br}.
At a lower density, the CO formation rate decreases ($\propto {\rm density}^2$) faster than the photodissociation rate ($\propto {\rm density}$).
Hence, a higher column density ($A_{\rm V}>$1) is required for the self-shielding of CO.
It is difficult to achieve such a condition and to maintain CO outside molecular clouds.

\citet{Roman-Duval:2016aa} showed that about 25\% of total molecular gas mass in the MW
is detected in CO, but not in $^{13}$CO.
They called it the ``diffuse" component.
\citet{Goldsmith:2008aa} performed a sensible analysis of the CO-bright, $^{13}$CO-dark
outer layer of the Taurus molecular cloud.
They found a similar mass fraction (37\%) in this layer
and derived the densities of $10^{2}$-$10^{3}\,\rm cm^{-3}$.
Hence, the CO-emitting ``diffuse" component across the MW is 
likely the cloud outer layers whose densities are as high as the average density of molecular clouds.

Therefore, when CO(1-0) emission is detected,
the majority is most likely emitted from molecular clouds,
even when individual clouds are spatially unresolved.
We note again that in this paper we use the term ``molecular clouds"
for all molecular gas concentrations  with the average cloud density.
We do not discuss how much of this gas is gravitationally-bound
\citep{Heyer:2001aa, Sawada:2012aa, Evans:2021aa},
as it is beyond the scope of this paper.

\subsection{Molecular Clouds and Their Spatial Variations} \label{sec:gmctemp}

The integrated intensity $\Ico$ map (Figure \ref{fig:combmom0}) could portray multiple overlapping clouds along any given line of sight,
particularly in the central region.
On the other hand, the individual peaks in the $\Tp$ map (Figure \ref{fig:combmom8})
likely represent those of individual clouds.
The CO(1-0) emission is typically optically-thick, and
when a cloud is spatially resolved with the $40\pc$ beam
(i.e., the beam filling factor is $f=1$),
the observed brightness temperature $\Tb$ (and $\Tp$) directly reflects the kinetic temperature
$\Tk$ of the bulk molecular gas in the cloud (see Figure \ref{fig:tbtkin}).
The prominent molecular structures (e.g., offset ridges, spiral arms,
large interarm structures) are wider than the typical diameter of molecular clouds,
and their areas are likely filled with clouds.
Hence, we assume $f=1$ within those structures.

Figure \ref{fig:combmom8} shows that
$\Tp$ decreases with galactic radius, indicating that
molecular clouds on average are the warmest in the central region and become cooler toward the outer disk.
The highest $\Tp$ is located in the central region (see also Figure \ref{fig:pdfs}e),
with a maximum of $\sim 16$-$18\K$, corresponding to $\Tk\sim 19$-$21\K$.
This is comparable to the kinetic temperatures
of the clouds in the Galactic center \citep{Oka:2001qt}
and is twice as warm as that of typical clouds in the MW disk
\citep[Table \ref{tab:gmcmw}; ][]{Scoville:1987vo, Sawada:2012aa}.
Beyond the central region,
the high brightness regions ($\Tp \sim 6$-$10\K$, or $\Tk \sim 9$-$13\K$)
are localized around the bar and spiral arms in the inner disk (see also Figure \ref{fig:pdfs}e).
Similarly high $\Tp$ values are rarely seen in the outer disk,
where the highest values are $\Tp \sim 4$-$8\K$ ($\Tk \sim 7$-$11\K$).

Radial variations are also seen in $\sigma_{v}$.
The $\sigma_{v}$ is measured at a cloud-scale resolution (40~pc)
and likely traces the dispersion within individual molecular clouds,
except in the central and bar regions where the line-of-sight overlap
of multiple clouds may be an issue.
The typical $\sigma_{v}$ decreases from $\sim 6.0\kmps$ in the inner disk
to $\sim 4.8\kmps$ in the outer disk.
Hence, the properties of molecular clouds change with galactic radius,
and $\Tk$ and $\sigma_{v}$ decrease radially.
We note that the observed $\sigma_{v}$ is supersonic
even at the edge of the disk (the sound speed is $\sim0.24\kmps$ at 10~K).

Azimuthal variations in $\Tp$ ($\Tb$) are also clear in Figure  \ref{fig:combmom8}.
In the inner disk,
significant interarm structures (e.g., filaments)
exhibit the temperatures as low as $\Tp\lesssim 3\K$.
It may be safe to assume $f=1$ in these structures
as they maintain a high surface density over a width of $\gtrsim 100\pc$ (Sections \ref{sec:innerdisk} and \ref{sec:outerdisk}).
If $f=1$, $\Tk \sim 6\K$ in those interarm structures,
which is lower than $\Tk \sim 9$-$13\K$ in the main spiral arms in the inner disk.
Such an arm-intearm variation in molecular gas temperature has also been reported 
by a CO 2-1/1-0 line ratio analysis of M83 at a lower resolution \citep{Koda:2020aa}.

\subsection{Extended CO Emission} \label{sec:extendedCO}

The extended CO(1-0) emission is present
in and around the bar and spiral arms
(Figures \ref{fig:combmom0} and \ref{fig:combmom0alternative}).
As discussed in Section \ref{sec:COexcitation},
it is unlikely that the majority of this CO emission comes from
diffuse molecular gas outside molecular clouds.
Instead, it is likely from small, unresolved clouds.

Figure \ref{fig:combmom8}, as well as Figures \ref{fig:combmom0} and \ref{fig:combmom0alternative}, show that
the low surface brightness emission ($\Tp \lesssim 0.5\K$) is ubiquitous.
It is associated with, but more extended than, the narrow molecular structures (bar, spiral arms, interarm structures).
If this extended emission consists of unresolved clouds,
the beam filling factor is $f<1$.
As a thought experiment,
if we arbitrarily assume their kinematic temperature to be $\Tk\sim 5\K$
(i.e., $1\K$ lower than those of the significant interarm structures,
but higher than $T_{\rm CMB}$),
Figure \ref{fig:tbtkin} gives a corresponding value of $\Tb/f\sim 2\K$
\citep[roughly consistent with the most frequent $\Tb$ value in the Taurus cloud in the MW from Figure 4 of ][ especially when the large uncertainty of their main beam efficiency for extended source is taken into account]{Narayanan:2008aa}.
Thus, for the observed $\Tb = \Tp \sim 0.5\K$,
the filling factor in our $40\pc$ beam would be $f\sim 1/4$.
If the emission is from a single cloud, its diameter is $D=20\pc$,
and the gas mass is $\Mgas=1.4\times 10^4~\Msun$ within a $5\kmps$ channel width.
Such molecular clouds are abundant in the MW disk.
While it cannot be proven at our spatial resolution, this cloud-based explanation
of the extended CO emission seems reasonable.

\section{Dynamical Organizations of Molecular Structures} \label{sec:dynamicalorigin}

Molecular structures in M83 are often $\gtrsim 1\kpc$ in length and 100-200~pc in width,
having masses of $10^7\Msun$.
They are ubiquitous at various radii across the disk even in the interarm regions
(Sections \ref{sec:innerdisk} and \ref{sec:outerdisk}).
Their diversity, at first glance, appears to preclude a unified scenario for their formation.
However, with a simple thought experiment (Sections \ref{sec:disktimescales} and  \ref{sec:formtimescale}), 
we suggest that galactic dynamics around spiral arms play a determinant role,
even for the interarm structures.
The majority of clouds should have formed in stellar spiral arms,
moved out from the parental arms without being dispersed,
or remained intact after the stellar arms were dissolved.
This dynamically-driven scenario
suggests that the molecular structures move over substantial distances
across the disk, which takes time and indicates the long lifetimes of these structures and constituent molecules once formed (Section \ref{sec:dynamicevolution}).

In this section, we illustrate how the molecular structures can form and evolve
with basic considerations regarding their formation timescale.
We lay emphasis on the numerous filamentary structures detected in the interarm regions.
They often extend over $\gtrsim1\kpc$ in length and 100-200~pc in width
with a surface density enhancement of $\gtrsim 10$
in contrast to the surrounding regions (Figures \ref{fig:combmom0} and \ref{fig:combmom0alternative}).

\subsection{Disk Dynamical Timescales} \label{sec:disktimescales}

For reference,
we calculate two dynamical timescales using the parameters of M83 (Table \ref{tab:m83}).

The rotation timescale of the disk is
\begin{equation}
    t_{\rm rot, disk}\sim 106\Myr \left( \frac{R}{3\kpc} \right) \left( \frac{V_{\rm rot}}{174 \kmps} \right)^{-1}.\label{eq:trotdisk}
\end{equation}

The rotation of the bar and spiral pattern is also important.
However, a growing number of studies support that spiral arms are dynamically-varying, transient structures rather than the static pattern postulated by the classic density-wave theory \citep{DOnghia:2013aa, Baba:2013vl,Dobbs:2014aa, Sellwood:2021wi}.
At least, the bar pattern must be static since otherwise it cannot maintain its straight shape.
Using a bar pattern speed of $\Omega_{\rm p} \sim 57.4\kmps \, \kpc^{-1}$
\citep{Hirota:2014wd}, the rotation timescale of the bar is
\begin{equation}
    t_{\rm rot, pattern} \sim 107\Myr \left( \frac{\Omega_{\rm p}}{57.4\kmps \, \kpc^{-1}} \right)^{-1}.\label{eq:trotbar}
\end{equation}

\subsection{Gas Assembly and Molecular Structures} \label{sec:formtimescale}

The long lengths ($\gtrsim 1\kpc$) of the narrow molecular structures (100-200~pc) already
suggest the importance of large-scale galactic dynamics in their formation.
We consider a case that
a long and narrow structure is formed
by a converging gas flow compressing a sheet of ambient gas in one direction (like forming a wrinkle on a sheet).
The surface densities and widths of the ambient gas (initial condition)
and of the molecular structure (final) are $\Sigma_{\rm i}$, $l_{\rm i}$, $\Sigma_{\rm f}$, 
and $l_{\rm f}$, respectively.
The mass conservation gives the width of the initial region,
\begin{equation}
l_{\rm i} = \left( \frac{\Sigma_{\rm f}}{\Sigma_{\rm i}} \right) l_{\rm f}.
\end{equation}

In the process, the gas has to move over a distance of $(l_{\rm i}-l_{\rm f})$.
We adopt a simplistic assumption of a coherent converging flow
at a constant velocity width of $\Delta v$
(i.e., \textit{all} gas over an area of width $l_{\rm i}$ is moving
in the same directions toward a narrow structure).
The formation timescale of the structure is
\begin{equation}
    t_{\rm form} = \frac{l_{\rm i}-l_{\rm f}}{\Delta  v}=\frac{l_{\rm f}}{\Delta v} \left( \frac{\Sigma_{\rm f}}{\Sigma_{\rm i}}-1 \right) \approx \frac{l_{\rm f}}{\Delta  v} \frac{\Sigma_{\rm f}}{\Sigma_{\rm i}}.
\end{equation}
The last approximation is for a large contrast between the formed structure
and ambient gas ($\Sigma_{\rm f}/\Sigma_{\rm i} \gg 1$).

Here we focus on 
how long it takes the filamentary molecular structures to form in interarm regions,
and whether such formation is possible across the disk.
In order to build up the mass of a molecular structure
of width, $l_{\rm f}\sim 100\pc$, and contrast, $\Sigma_{\rm f}/\Sigma_{\rm i} \sim 10$,
the initial width has to be
\begin{equation}
l_{\rm i} \sim 1\kpc \left( \frac{\Sigma_{\rm f}/\Sigma_{\rm i}}{10} \right) \left( \frac{l_{\rm f}}{100\pc} \right).
\end{equation}
The gas over a 1~kpc area must be converging coherently.
Obviously, a wider structure (e.g., $l_{\rm f} \sim 200\pc$) requires
an even larger initial area to sweep up the mass ($l_{\rm i} \sim 2\kpc$).
Such a coherent converging flow over such a large area is  unlikely in the interarm regions.

A detailed analysis of the velocity field is necessary to derive $\Delta v$,
which is beyond the scope of this paper.
The $\Delta v$ is not represented by the velocity dispersion since it does not cause a coherent convergence.
The large-scale converging flow should be due to non-circular orbital motions.

Here, we simply adopt $\Delta v=10\kmps$ as a fiducial value in the interarm regions,
surmised from a measurement in M51, a galaxy with more prominent spiral arms.
\citet{Meidt:2013to} derived mass-weighted azimuthally-averaged non-circular velocities
of about 5-25$\kmps$ in M51 (their Figure 2).
This azimuthal average includes the spiral arms and interarm regions,
and generally, the spiral arms have larger non-circular velocities than the interarm regions.
Hence, a typical velocity in the interarm regions is likely on the smaller side of, or less than, this range.
Additionally, when neighboring gas travels together on the same non-circular motions,
their convergence velocity $\Delta v$ should be smaller than the non-circular velocities themselves.
$\Delta v=10\kmps$ may be on a large side within the expected range in the interarm regions.

Hence the formation of the filamentary structures in interarm regions, if they form in-situ, requires
\begin{equation}
t_{\rm form} \sim 98\Myr \left( \frac{l_{\rm f}}{100\pc} \right) \left( \frac{\Delta v}{10\kmps} \right)^{-1} \left( \frac{\Sigma_{\rm f}/\Sigma_{\rm i}}{10} \right).\label{eq:tfrm}
\end{equation}
This is as long as the rotation timescales of the disk and pattern (eqs. \ref{eq:trotdisk} and \ref{eq:trotbar}).
This estimation is crude, but illustrates that their formation in the interarm regions would take substantial times
with respect to the dynamical timescales.

Together with the required condition for a coherent converging flow ($\gtrsim 10\kmps$)
over a large scale ($\gtrsim 1\kpc$),
the required long timescale suggests that the formation of the filamentary structures in interarm regions is unlikely.
This assessment is for general cases and to explain the ubiquity of the structures at various radii over the disk.
Of course, some exceptional cases may still be found,
but they are not applicable in general.

Equation (\ref{eq:tfrm}) shows that, for quick formation ($\sim 30\Myr$),
the gas has to move coherently at a fast speed ($\gtrsim 30\kmps$)
into the observed filamentary structures,
independent of the cause of the converging motion.
In addition, these structures are massive ($\gtrsim 10^7\Msun$; Section \ref{sec:submass}),
and this amount of gas has to assemble from gas distributed over a $>1\kpc$ region.
Observationally, it is very rare to find such massive, large-scale coherent gas motions outside spiral arms
[unless the \textit{majority} of the molecular gas is hidden in the CO-dark phase \citep[see ][]{Pringle:2001zi} --
however, its fraction is only $\lesssim 30\%$ within the solar radius of the MW \citep{Pineda:2013lr}].
This observational constraint of a fast, coherent flow velocity
of a large amount of gas must be satisfied by
any potential mechanism to be considered for the formation of the filamentary structures.

In spiral arms, the mass naturally converges due to the large-scale gravitational potential.
With a consideration similar to the above,
we could set $\Delta v \gtrsim 30\kmps$ for spiral arms, based again on \citet{Meidt:2013to}.
The formation timescale is shorter in the spiral arms: $t_{\rm form}\lesssim 30\Myr$ for $\Delta v \gtrsim 30\kmps$.
Therefore, the filamentary structures can form in spiral arms.

Stellar feedback could also push the gas and form structures in the interarm regions.
However, the feedback-driven models so far explained only less massive structures
\citep[$\sim 10^6\Msun$; ][]{Smith:2020aa, Tress:2021aa, Kim:2022ab}.
Even for the small masses, these models include spiral arm potentials
and assemble the gas primarily by the potentials.
The role of feedback in the assembly process appears secondary.

\subsection{Dynamically-Driven Evolution} \label{sec:dynamicevolution}

The above consideration suggests that 
the interarm molecular structures should have formed in stellar spiral arms.
They would have traversed across the widths of the spiral arms and flowed into the interarm regions
as suggested by the density-wave theory \citep{Fujimoto:1968tm, Fujimoto:1968wl, Roberts:1969ct},
or remained after the stellar spiral arms have dissolved
as predicted by the swing amplification theory \citep{Toomre:1981uv, DOnghia:2013aa, Baba:2013vl}.
In either case,
the formation and evolution of the diverse molecular structures are
driven mainly by large-scale galactic dynamics.

The classic density-wave theory postulates steady, long-lived, stellar spiral arms,
which can accumulate coherent, large-scale molecular spiral arms.
Some parts of the molecular arms become large molecular concentrations
\citep[massive molecular clouds; ][]{Vogel:1988tp, Aalto:1999wk},
which can be stretched by the shearing force along the spiral arms
and by the differential rotation after the spiral arm passages.
They naturally extend to spurs/feathers in the interarm regions
(\citealt{Koda:2009wd}; see also \citealt{Corder:2008aa, Schinnerer:2017aa}).
In the inner disk of M83, the molecular gas spiral arms appear relatively focused as narrow ridges (about 600~pc widths).
The spurs/feathers are resolved as chains of molecular clouds \citep{Koda:2009wd}, and often rooted to, and extending out of, the main spiral arms.
Notably, they are found mostly in the inner disk,
predominantly on the convex sides of the spiral arms (Section \ref{sec:innerdisk})
as expected in the density-wave type of gas flow.

The swing amplification model predicts transient, short-lived stellar spiral arms,
where the gas also accumulates, dissipates, and becomes denser to form molecular structures
\citep{Baba:2015uc, Baba:2015tt}.
The stellar spiral structures are temporary density enhancements, but self-perpetuate themselves by forming subsequent spiral arms
nearby \citep{DOnghia:2013aa}.
The temporary nature results in less coherent gas structures \citep{Baba:2015uc, Baba:2015tt},
which can be left after the stellar spiral arms dissolve.
In the outer disk of M83, the molecular spiral arms continue from the inner disk,
but are broader, less spatially coherent, and appear flocculent.

\citet{Baba:2015uc} suggested a gradual radial transition in a barred spiral galaxy,
from the density-wave picture in the inner disk to the swing amplification picture in the outer disk.
In their model, even the inner spiral arms are triggered by the swing amplification, but live longer due to the gravitational influence of the static bar potential,
and behave like steady density waves for some period of time
(i.e., a good fraction of the disk rotation timescale - eq. \ref{eq:trotdisk}).
The observed transition in M83 from the coherent to less-coherent molecular spiral arms from the inner to outer disk may fit to this transition picture.

The bar region also shows filamentary molecular structures.
Some of these features run into the offset ridges from their upstream sides, 
assuming that the disk is rotating in the clockwise direction (Figure \ref{fig:combmom0}).
They could also be sheared-off remnants that traveled from the previous offset ridge.
Given a relatively short crossing timescale of $\sim 20$-$80\Myr$ \citep[][ their Figure 17]{Hirota:2014wd},
their remnants can reach the next offset ridge before they are completely broken apart.
Such a formation mechanism of stretched filamentary structures has been discussed
with simple orbit models in bar potentials \citep{Koda:2006aa, Hirota:2014wd}.

The discussions here are based only on a basic estimation of
their formation timescale and on their morphologies.
It is not surprising that some exceptions exist.
For example, isolated molecular clouds in the interarm regions,
which appear as dots in Figures \ref{fig:combmom0}-\ref{fig:combmom8},
could have formed in local density fluctuations,
rather than a result of galactic dynamics.

\subsection{Implications on Cloud Lifetimes} \label{sec:cloudlifetime}
The discussion above suggests that the large molecular structures survive
through the dynamical processes in the galaxy
and have lifetimes of an order of $\sim 100$ Myr.
Since the CO(1-0) emission is primarily from molecular clouds
(see Section \ref{sec:COexcitation}),
it implicates that the lifetimes of the embedded molecules and
molecular clouds are also as long.

Here, as \textit{a} definition of molecular cloud lifetimes, we adopt
the full duration that the gas is molecular in molecular clouds,
and hence, is in a prerequisite condition for potential star formation.
It has been suggested that
molecular clouds evolve through coagulation and fragmentation
\citep{Scoville:1979lg, Vogel:1988tp, Koda:2009wd, Dobbs:2013aa},
and that some parts within clouds (e.g., dense cores) could be dispersed
by stellar feedback \citep[e.g., ][]{Elmegreen:2007aa}.
Those processes might be acting within the observed molecular structures.
Some authors defined cloud lifetimes as
a short branch period between a coagulation and fragmentation
\citep[e. g., ][]{Kruijssen:2019aa, Chevance:2020aa, Kim:2022aa},
which is a correct definition in its own way.
However, most cloud studies stem from interests in star formation,
and it seems more relevant to adopt the full duration that the gas can potentially form stars as cloud lifetimes.

\section{Summary and Conclusions}

We present an ALMA imaging of molecular gas
across the full star-forming disk of the barred spiral galaxy M83 in CO(1-0).
The joint deconvolution of the data from the ALMA 12m, 7m, and Total Power arrays
was performed with the MIRIAD and TP2VIS packages.
The mass sensitivity of $10^4~\Msun$ is sufficient to detect the most abundant
population of molecular clouds with masses $\lesssim 10^5~\Msun$.
The spatial resolution of 40~pc is the typical physical diameter
of clouds in the MW ($D\sim 40~\pc$).
Therefore, this case study of one galaxy
is complementary to the large survey of
nearby galaxies in the excited CO(2-1) transition 
\citep{Leroy:2021aa, Leroy:2021ab},
which mainly detects a population of massive molecular clouds of $\gtrsim 10^5~\Msun$
\citep{Rosolowsky:2021aa} at a spatial resolution of 45-120~pc,
high enough to separate those clouds, but not to resolve the typical cloud diameter
\citep[][]{Sun:2018aa, Sun:2020tt}.

The molecular gas distribution shows coherent large-scale structures in the inner part,
including the central gas concentration, offset ridges along the bar,
and prominent molecular spiral arms in the inner disk.
In the outer disk,
the molecular spiral arms are still present,
but appear less coherent and flocculent.
Massive filamentary gas concentrations are present both in the spiral arms
and interarm regions in the inner and outer disks, as well as in the bar region.
All of these structures embed molecular clouds and appear as chains of clouds.
Many of the interarm structures host little or no star formation traced by H$\alpha$ emission.
Unresolved, extended CO emission is also detected around these structures.

The data show the radial and azimuthal variations in molecular gas properties,
from the galactic center, bar, and inner to outer disks,
and between the spiral arms and interarm regions.
The local properties,
such as brightness temperature, velocity dispersion,
integrated intensity, and hence, surface gas density, decrease outward.
They are higher in the spiral arms and lower in the interarm regions.
Given that the spatial resolution is comparable to the clouds' typical diameters,
these radial and azimuthal variations can be attributed to the variations of
clouds' internal properties.
The majority of the detected CO(1-0) emission,
including the unresolved CO emission, is most likely from molecular clouds
(cloud-like gas concentrations).

We describe a scenario in which the ubiquitous large molecular structures,
especially the ones in the interarm regions, can form and evolve.
It would take too long to assemble their huge masses without 
coherent converging gas flows over very large areas.
Such flows are expected around spiral arms, but not in the interarm regions.
The interarm structures are therefore unlikely to form in-situ.
Instead, we suggest that they assemble within the stellar spiral potential,
and are either expelled into the interarm regions from the spiral arms,
or are left behind after the stellar arms dissolve.
This indicates that the molecular structures, and embedded molecular clouds,
survive through these dynamical processes and have lifetimes of the same order as the rotation timescale
of the disk ($\gtrsim 100$ Myr).
The overall distribution of the molecular structures appears to be consistent
with the gradual radial transition
from the spiral density-wave picture in the inner disk
to the swing amplification picture in the outer disk.

This picture of dynamically-driven molecular gas evolution is suggested 
to explain the majority of the molecular structures and clouds observed in M83.
The discussions are based only on a basic estimation of
their formation timescale and on their morphologies.
It is not surprising that some exceptions exist.
We expect that subsequent studies and future observations will
confirm or modify this picture.

\begin{acknowledgments}
We thank the anonymous referee for useful comments.
We thank all staff members at the Joint ALMA Observatory (JAO) and North American ALMA Science Center (NAASC), who helped us customize the observation strategy and made the data reduction possible.
JK also thanks Peter Teuben for his help with MIRIAD and TP2VIS.
This paper makes use of the following ALMA data: ADS/JAO.ALMA\#2017.1.00079.S.
ALMA is a partnership of ESO (representing its member states), NSF (USA) and NINS (Japan), together with NRC (Canada), MOST and ASIAA (Taiwan), and KASI (Republic of Korea), in cooperation with the Republic of Chile. The Joint ALMA Observatory is operated by ESO, AUI/NRAO and NAOJ.
The National Radio Astronomy Observatory is a facility of the National Science Foundation operated under cooperative agreement by Associated Universities, Inc..
This research has made use of the NASA/IPAC Extragalactic Database (NED), which is operated by the Jet Propulsion Laboratory, California Institute of Technology, under contract with the National Aeronautics and Space Administration.
JK acknowledges support from NSF through grants AST-1812847 and AST-2006600,
and from NAASC through the ALMA Development Study program to develop TP2VIS.
FE is supported by JSPS KAKENHI grant No. 17K14259.
AGdP has been partly supported by grant RTI2018-096188-B-I00 funded by MCIN/AEI/10.13039/501100011033.
LCH was supported by the National Science Foundation of China (11721303, 11991052, 12011540375, 12233001) and the China Manned Space Project (CMS-CSST-2021-A04, CMS-CSST-2021-A06).
\end{acknowledgments}

\vspace{5mm}

\facilities{ALMA}
\software{CASA, MIRIAD, DS9, BBarolo}

\bibliographystyle{aasjournal}

\end{document}